\documentclass{article}

\usepackage{arxiv}
\usepackage[T1]{fontenc}
\usepackage[utf8]{inputenc}
\usepackage{textcomp}
\usepackage[american]{babel}
\usepackage{microtype}
\usepackage{amsmath}
\usepackage{amssymb}
\usepackage{graphicx}
\usepackage{booktabs}
\usepackage{multirow}
\usepackage{siunitx}
\usepackage{enumitem}
\usepackage{float}
\usepackage{caption}
\usepackage{csquotes}
\usepackage[backend=biber,style=apa,sortcites=true]{biblatex}
\DeclareLanguageMapping{american}{american-apa}
\usepackage[hidelinks]{hyperref}

\graphicspath{{figures/}}

\newcommand{\occunit}{\ensuremath{\mathrm{h}^{-1}\mathrm{m}^{-2}}}

\title{Do Waders, Swimmers, and Divers Exist? A GPS-Based Pilot Study of Site-Dependent Visitor Movement in Theme Parks}

\author{%
  Dane M. Utley \\
  Department of Civil and Environmental Engineering\\
  Princeton University, Princeton, NJ 08544
  \And
  J\"urgen Hackl\thanks{Corresponding author: \texttt{hackl@princeton.edu},
    ORCID \href{https://orcid.org/0000-0002-8849-5751}{0000-0002-8849-5751}.} \\
  Department of Civil and Environmental Engineering\\
  Princeton University, Princeton, NJ 08544\\
  \texttt{hackl@princeton.edu}
}
\date{\today}

\renewcommand{\shorttitle}{Waders, Swimmers, and Divers in Theme Parks}
\renewcommand{\headeright}{}   
\renewcommand{\undertitle}{}   

\hypersetup{
  pdftitle={Do Waders, Swimmers, and Divers Exist? A GPS-Based Pilot Study of Site-Dependent Visitor Movement in Theme Parks},
  pdfauthor={Dane M. Utley, Juergen Hackl},
  pdfkeywords={GPS tracking, visitor behavior, theme parks, trajectory clustering, behavioral segmentation, agent-based pedestrian simulation},
}

\begin{document}
\maketitle

\begin{abstract}
Theme-park master planners informally sort guests into behavioral types (waders, who
sightsee aimlessly; swimmers, who plan moderately; and divers, who maximize attractions) to guide layout
decisions and to seed pedestrian simulations, yet the typology is rarely grounded in observed movement.
To ground it, GPS tracks and an exit survey are collected from 23 recruited
guests, yielding 26 park visits across three Southern California parks (Knott's Berry Farm, $n=15$;
Disneyland, $n=7$; Disney California Adventure, $n=4$). Each visit is summarized by eight manually
annotated movement features; visits are clustered per park and the partitions assessed with a
multi-criteria protocol: internal validity indices, the gap statistic, restart and subsampling
stability, and agreement across clustering algorithms. Geometric separation is modest
everywhere (no silhouette $\ge 0.35$), yet at Knott's the three-group partition is highly reproducible
($k$-means and Ward agree exactly; consensus co-assignment 0.79 within versus 0.09 between), and its groups
differ on behavioral rates (attractions per hour, $p=0.004$). Reproducibility without sharp
geometric separation indicates a behavioral continuum rather than discrete types. The exit survey does
not distinguish the clusters, so objective movement and self-report diverge. Feature
directionality reverses across parks for three of eight features, so parameters cannot be universalized.
The observed heatmaps carry a monotone type signature, a thrill-versus-commercial occupancy gradient
that rises from waders to divers, and an agent-based test localizes its source: agent populations
differing only in measured movement parameters fail to separate the types, as does a homogeneous
baseline, whereas a model adding a priori type-specific destination preference and visit sequencing
reproduces the gradient's direction and ordering. Type-specific space
use therefore arises from destination choice more than from how visitors walk. Visitor-behavior
parameters are site-specific, and a reproducible, critically validated segmentation pipeline offers a
route to calibrating them per park.
\end{abstract}


\begin{refsection}

\section{Introduction}

On a single operating day a 100-acre theme park such as Disneyland moves tens of thousands of
guests several miles each on foot \parencite{ThemedEntertainmentAssociation2024TEA}. Sustaining the seamless visual storytelling that
defines these venues depends on keeping that pedestrian traffic flowing, so understanding and simulating
guest movement is a central task of theme-park master planning, both before a park is built and
throughout its operating life \parencite{Li1995study}. A planner who can anticipate where crowds will form
can shape sightlines, place retail and dining, size walkways, and stage attractions to spread demand.
The difficulty is that there is no single ``guest'' whose movement can be planned for. Visitors arrive
with different goals, energy, and tolerance for crowds, and they pursue those goals in different ways
across the same physical space, so the spatial demand a park experiences is the superposition of several
distinct movement styles rather than the behavior of one average visitor.

The stakes of getting this superposition right are operational and economic. Walkway width, queue
placement, retail frontage, show capacity, and the staging of headline attractions are all sized against
anticipated pedestrian load, and the cost of error is asymmetric: under-provisioned circulation produces
congestion that degrades the guest experience and, at the extreme, raises safety and egress concerns,
while over-provisioned circulation wastes land that in a theme park is exceptionally expensive per square
foot. Because a park cannot be re-poured once built, operators turn to pedestrian simulation to test
layouts before construction and to evaluate operational changes afterward, and the realism of those
simulations depends on how faithfully the agent population reproduces the mix of real movement styles. A
simulation seeded with a single average agent will systematically misplace demand, smearing it evenly
where real guests concentrate at thrill nodes or linger in retail cores; a simulation seeded with the
wrong proportions or the wrong per-type parameters will misplace it differently. The behavioral
parameters that distinguish visitor types are therefore not a modeling nicety but a determinant of the
predictions on which capital and safety decisions rest, which is what makes their empirical grounding
worth pursuing.

Industry practice manages this heterogeneity with a swimming-pool metaphor that sorts guests into three
types. \textcite{Younger2016Theme} describes \emph{waders} as those who ``just want to walk around and see
pretty pictures,'' \emph{divers} as those who ``want to know every shred of information that exists,''
and \emph{swimmers} as the bridging case. Translated into movement, waders are expected to wander
slowly, take scenic detours, and be drawn to incidental sights; divers to move quickly and directly so
as to ride as many attractions as possible; and swimmers to blend the two. The vocabulary is common
among designers and is used not only for attraction storytelling but, more relevant here, to seed the
digital pedestrian simulations on which operators increasingly rely for layout, safety, and revenue
decisions \parencite{Rajaram2003Flow}.

Yet the typology rests on expert judgment rather than measurement. To our knowledge, the proportions of
each type, and the movement parameters that distinguish them, have never been grounded in observed guest
trajectories. Existing academic theme-park models either optimize a single
named park from proprietary operational data \parencite{Ahmadi1997Managing,Rajaram2003Flow} or treat guests as
homogeneous agents that choose destinations by popularity and distance
\parencite{Solmaz2015Mobility,Vukadinovic2011simple}, leaving unmodeled exactly the behavioral heterogeneity the
wader/swimmer/diver framing presupposes. This gap matters in three distinct ways that prior work has not
separated. First, the categories may not correspond to any recoverable structure in measured movement at
all. Second, even if a partition is recoverable, it may be a fragile artifact of a particular algorithm
or sample rather than a reproducible grouping. Third, even a reproducible partition may carry different
behavioral meaning at different parks, so that parameter values calibrated at one venue mislead at
another. A credible empirical treatment of the typology must address all three questions, and it must do
so with validation that does not simply restate the assumptions used to construct the groups.

\paragraph{Research questions.} We ask: (RQ1) Can the wader/swimmer/diver typology be recovered from
GPS-derived movement features, and is the resulting partition reproducible across algorithms and
resamples rather than a single-method artifact? (RQ2) Do the recovered groups differ on theoretically
motivated movement features, and do they agree with an independent exit survey? (RQ3) Does the
behavioral meaning of each feature transfer across structurally different parks, or is it
park-dependent? (RQ4) Do cluster-derived parameters, when used to drive a microscopic pedestrian model,
reproduce observed spatial patterns, and do type-specific parameters add value over a homogeneous
baseline?

\paragraph{Contribution and scope.} This is an exploratory, park-specific pilot, not a confirmatory or
generalizable study, and we state that limit at the outset rather than burying it. We make no claim to a
representative sample, we report uncertainty rather than confirmatory hypothesis tests, and we do not
validate parameters in an operational deployment. Within those bounds the study makes four
contributions. First, it provides a reproducible pipeline that turns raw GPS tracks and a short exit
survey into eight interpretable movement features and a per-park segmentation, together with a
multi-criteria validation protocol (internal indices, gap statistic, restart and subsampling stability,
and cross-algorithm agreement) that distinguishes the three questions above. Second, it shows that at
the best-sampled park the typology is recoverable and highly reproducible even though the clusters are
not geometrically sharp, which we interpret as evidence of a behavioral continuum rather than discrete
types, and it shows that the partition is only weakly and non-significantly related to self-report.
Third, it quantifies the central finding that feature directionality is park-dependent, with a subset of
features acting as robust cross-park discriminators and others reversing sign. Fourth, it uses an
agent-based test to localize where the type signature lives: locomotion parameters alone do not reproduce
the observed thrill-versus-commercial occupancy gradient, but an a priori destination-choice and visit-sequencing
model does, identifying destination selection rather than walking mechanics as the operative mechanism. The strongest evidence comes from
Knott's Berry Farm ($n=15$); the two Disney samples ($n=7$, $n=4$) are too small for coherent clustering
and are treated as exploratory throughout. All data-processing and simulation code is released so that
the pipeline can be reused and the limited results scrutinized.

\section{Conceptual framework and expectations}\label{sec:framework}

Tourism research has long debated whether visitors fall into discrete behavioral types or vary along
continua. Typological traditions, from \textcite{Cohen1972Sociology} and \textcite{Plog1974Why} to
museum-visitor identities \parencite{Falk2016Identity}, posit a small number of recognizable kinds of visitor,
whereas movement studies repeatedly find that real itineraries grade into one another and that an
individual can shift style within a single visit \parencite{Mckercher2008Movement,Beeco2013GPS}. The
wader/swimmer/diver scheme is squarely a typological claim, and the central methodological question it
raises is therefore not merely ``do clusters exist'' but ``in what sense do they exist.'' We separate
three properties that the segmentation literature often conflates. \emph{Recoverability} asks whether a
partition matching the typology emerges from the data at all. \emph{Reproducibility} asks whether that
partition is stable to the choice of algorithm and to resampling, as opposed to a fragile artifact.
\emph{Separation} asks whether the groups are geometrically well isolated, as summarized by indices such
as the silhouette. These properties can dissociate: a behavioral continuum partitioned at $k=3$ can be
highly reproducible (the same participants are grouped together every time) while showing weak
separation (no gap between groups), and that combination is itself diagnostic of graded rather than
discrete behavior. We add a fourth property, \emph{external validity}, meaning agreement with
independent data such as a survey, and a fifth, \emph{utility}, meaning usefulness for the downstream
task of parametrizing a simulation. A segmentation can be reproducible and yet have low external
validity or limited utility, and distinguishing these outcomes is part of our contribution.

Against this framework we state the expectations that motivated the analysis, while emphasizing that the
study is exploratory and that we specified in advance only the cluster-coherence threshold described in the
Methods. We expected that (E1) a three-group partition would be recoverable and, if the typology is
real, reproducible across algorithms and resamples; (E2) the groups would differ on the movement
features that most directly encode diving versus wading, namely attraction-visitation rate, attraction
focus, and walking speed; (E3) the partition might nonetheless separate only weakly in feature space if
visitor behavior is graded; (E4) self-report would corroborate the groups, although prior work warns
that stated and revealed behavior diverge; (E5) the behavioral meaning of features would differ across
parks of different price, audience, and layout, so that a pooled model would fail and parameters would
not transfer; and (E6) cluster-derived parameters would improve a simulation relative to a homogeneous
baseline. As reported below, E1, E2, E3, E5, and E6 are supported and E4 is not; E6 holds only in
qualified form, the measured parameters improving a simulation once combined with type-specific
destination preference and sequencing, an outcome that is itself informative about where type-specific
spatial structure comes from.

\section{Related work}

\subsection{Theme-park and pedestrian simulation}
The earliest theme-park models were park-specific operational tools. \textcite{Ahmadi1997Managing} built a
flow-and-transition model of Six Flags Magic Mountain from a year of hourly attraction counts,
classifying guests by age band but never validating that the classification corresponded to distinct
movement. \textcite{Rajaram2003Flow} used participant diaries at Universal Studios Hollywood to optimize
retail flow, again from a single park's data. More recent work emphasizes transferable tools but, in
doing so, treats guests as homogeneous: \textcite{Solmaz2015Mobility} weight attraction popularity against
Euclidean distance, and \textcite{Vukadinovic2011simple} (ParkSim) route agents along shortest paths, both
conceding that constant-speed average agents miss the heterogeneity in how different guests traverse
pathways. \textcite{Cheng2013agentbased} demonstrate large-scale agent-based experience management in theme parks;
real-time reservation systems such as Lightning Lane increasingly reshape flow \parencite{Pache2024Theme};
and recent surveys of crowd simulation stress that injecting \emph{heterogeneous} behavior is the key
open problem \parencite{Khan2024Agentbased}. The agent substrate we target is microscopic pedestrian dynamics: the
social-force model \parencite{Helbing1995Social,Helbing2000Simulating}, velocity-obstacle methods such as ORCA
\parencite{VanDenBerg2011Reciprocal}, and cellular-automaton approaches \parencite{Schadschneider2001Cellular}. These
frameworks expose tunable per-agent parameters such as desired speed, goal commitment, and destination
choice, but provide no empirical basis for setting them by visitor type. Our work supplies that missing
link and, importantly, tests whether locomotion parameters alone are sufficient to reproduce
type-specific space use.

\subsection{GPS tracking and segmentation of visitors}
GPS tracking is by now an established method for analyzing the space-time behavior of tourists,
including in theme parks. \textcite{Birenboim2013Temporal} tracked guests at PortAventura and identified
distinct diurnal activity rhythms, establishing the feasibility and value of GPS in exactly this
setting, and \textcite{Huang2020Tourists} used handheld GPS at Ocean Park Hong Kong to identify three
spatial-temporal visitor clusters by density-center clustering, the closest published theme-park
analogue to the present study, though without the validation, survey triangulation, or cross-park
comparison we undertake here. In outdoor recreation, \textcite{Beeco2013GPS} tracked visitors and tried to separate
``wanderers'' from ``planners,'' the nearest antecedent of our typology, but found that a binary
split was insufficient and that intermediate behavior required a third category. That finding directly
motivates our three-way treatment and anticipates our own result that the boundaries between types are
soft. A growing body of work clusters GPS and related movement traces into a small number of visitor types in
national parks \parencite{Choe2023Identifying}, mountainous scenic areas \parencite{Liu2022Cluster}, and large-scale
events \parencite{Abkarian2022Characterizing}, and increasingly fuses several data sources to reconstruct movement at
destination scale \parencite{AbreuNovais2026Investigating}, typically recovering three to six behavioral groups that
differ in distance traveled, dwell time, and activity sequencing. At a broader level, tourist-movement modeling
\parencite{Lew2006Modeling,Mckercher2008Movement}, travel-persona segmentation \parencite{Park2010Travel}, and classic
visitor typologies \parencite{Plog1974Why,Cohen1972Sociology,Falk2016Identity} establish that segmenting visitors by
behavior is meaningful, while cautioning that an individual may shift type within a single visit. Our
study adds the comparative dimension that this literature largely lacks: rather than segmenting visitors
at one site, we ask whether the \emph{same} feature defines the same type across structurally different
parks, and we find that several do not.

It is worth situating our recovered structure against this body of work quantitatively. Prior GPS
clustering studies typically report a partition (three sequence-aligned types in a national park
\parencite{Choe2023Identifying}, several movement types in mountainous scenic areas \parencite{Liu2022Cluster}, six
activity-sequence groups at a large event \parencite{Abkarian2022Characterizing}) but seldom report the
separation, stability, or cross-algorithm diagnostics that would establish in what sense those
partitions exist. The number of recovered groups is in part a modeling choice tied to the application,
and our three-group partition is comparable in granularity to the smaller of these schemes. What
distinguishes the present study is less the partition than the validation: by reporting separation,
reproducibility, and algorithm agreement separately, and by testing the partition against an independent
survey and across multiple parks, we can say not only that a three-group structure can be drawn but how
trustworthy and how transferable it is, which is the information a planner actually needs before relying
on it.

\subsection{Clustering methods and validity}
We cluster with $k$-means \parencite{MacQueen1967methods}, the most widely used and interpretable partitioning
method, but we do not treat a single solution as ground truth. Because $k$-means requires the number of
clusters in advance and is sensitive to initialization, we assess each solution with internal validity
indices, namely the silhouette coefficient \parencite{Rousseeuw1987Silhouettes}, the Davies--Bouldin index
\parencite{Davies1979Cluster}, and the Caliński--Harabasz index \parencite{Calinski1974dendrite}, choose the number of
clusters with the gap statistic \parencite{Tibshirani2001Estimating}, and quantify reproducibility through both
restart stability and subsampling consensus clustering \parencite{Monti2003Consensus}. We additionally compare
the partition against agglomerative (Ward) and Gaussian-mixture solutions, because agreement across
algorithms with different inductive biases is strong evidence that a grouping is not a method artifact.
Density-based and trajectory-specific alternatives such as DBSCAN \parencite{Ester1996densitybased} and TRACLUS
\parencite{Lee2007Trajectory} (reviewed by \textcite{Yuan2017review}) are appropriate when clusters are non-convex or
when one clusters raw trajectories rather than engineered per-visit feature vectors. Because our features
are interpretable per-visit summaries chosen to map onto the industry typology, we retain $k$-means for
transparency while reporting its limitations candidly.

\section{Methods}

\subsection{Study sites}
The three parks span the structural contrast that motivates our park-dependence question. Knott's Berry
Farm, in Buena Park, California, is a regional, mixed-thrill park with comparatively low admission prices
and no-blackout annual passes, and it draws a largely local, repeat-visit audience. The Disneyland
Resort in Anaheim, comprising Disneyland and Disney California Adventure (DCA), sits at the opposite end:
admission and annual passes cost several times more and are reservation-gated
\parencite{disney-tickets,disney-magickey,knotts-tickets}, the audience is more destination-oriented, and the
environment is saturated with themed retail and immersive experiences that make spending and dwelling
part of the visit. Disneyland is the more heavily attended of the resort's two gates \parencite{ThemedEntertainmentAssociation2024TEA},
and its hub-and-spoke plan funnels guests through central commercial cores, whereas DCA follows a looser
waterfront layout. These differences in price, audience, theming intensity, and topology are precisely
the conditions under which we expect the behavioral meaning of a movement feature to shift from one park
to another.

\subsection{Study design and data collection}
Over three operating days in March 2026 we recruited 23 consenting adult guests at the entrances of
Knott's Berry Farm (two days), Disneyland, and DCA, under an approved human-subjects protocol
(\ifdefined\anonymous institutional review board details withheld for anonymous review\else Princeton
University IRB \#19130\fi; the recruitment script, consent form, and survey instrument appear in the
Supplementary Material). Each participant carried a Canmore GT-730FL-S USB logger that recorded position
once per second. The device does not transmit live location, a deliberate privacy-preserving choice that
meant data could be extracted only after the logger was returned at the end of the day. On return, each
participant completed a five-item exit survey covering prior-visit recency, itinerary planning,
attraction priorities, frequency of map consultation, and a self-described movement style (wandering,
in-between, or beelining), and received a \$15 gift card.

Four of the seven Disneyland participants park-hopped to DCA the same day; their DCA visits are analyzed
separately. One Knott's participant who stayed only \SI{45}{\minute} for lunch
and visited no attractions was excluded as an operational outlier. The resulting sample is 26
park visits: $n=15$ at Knott's, $n=7$ at Disneyland, and $n=4$ at DCA. The sample is small,
single-season, and non-representative, and we return to those threats in Section~\ref{sec:limits}.

The recruitment experience is itself a methodological lesson worth recording, because it constrains what
in-person GPS studies of theme parks can achieve. On-site recruitment at the park gate yielded both a
modest number of consenting participants per day and a sample skewed toward passholders, who proved more
willing than once-a-year visitors to pause at the threshold of an anticipated visit and to engage a
researcher. Families with children, a large share of real attendance, declined uniformly, plausibly
because parents arriving with excited children have little attention to spare. Park operators also limit
on-property research, which caps the number of collection days available. These frictions push toward
alternative designs for future work, in particular virtual or pre-arranged recruitment that reaches a
broader cross-section, including first-time and family visitors, and that decouples the consent
conversation from the high-anticipation moment of entry. We flag this not as an aside but as practical
guidance: the demographic composition of a GPS sample is shaped as much by where and how recruitment
happens as by the underlying visitor population, and study designs should anticipate and counteract that
bias rather than discover it after the fact.

\subsection{Trajectory cleaning and feature extraction}
\begin{figure}[t]
\centering
\includegraphics[width=1\linewidth]{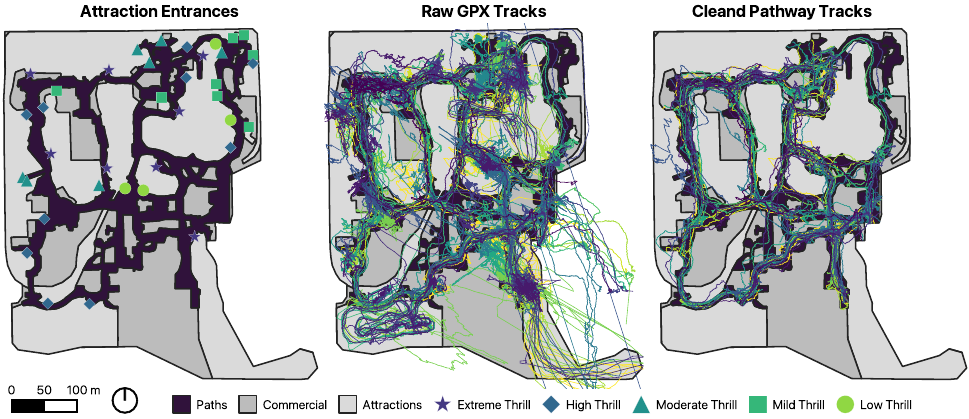}
\caption{Trajectory cleaning at Knott's Berry Farm. Left: the park layout, showing the pathway network,
commercial zones, attraction envelopes, and ride access points. Middle: all raw imported GPX trajectories
across participants, before cleaning. Right: the same trajectories after a single analyst removed time
spent inside attraction, retail, and restroom envelopes and excised corrupted segments, leaving the
pathway-only tracks that are the input to the movement features.}
\label{fig:cleaning}
\end{figure}

Raw GPX trips were merged and reprojected in a geographic information system (QGIS). Because consumer GPS
is not accurate enough to clip reliably to pathway polygons, a single analyst parsed each track
chronologically, classifying time as pathway travel versus dwell inside an attraction, retail, or
restroom envelope, and logging every point-of-interest (POI) visit with its name, entry and exit times,
and walked versus shortest-path distance. Segments showing physically impossible jumps, caused by indoor
signal loss or ride dynamics, were removed; for some participants this cost up to two hours of an
eight-hour day. For two participants
whose tracks were corrupted at the very start of the visit (one at DCA, one at Knott's), time to first
attraction could not be measured and was set to the park mean, a neutral imputation that does not push
the participant toward either extreme of that feature. Manual coding by one analyst is a limitation: we
did not compute inter-rater reliability, and we recommend that future work have a second coder
re-annotate a track subsample so that agreement can be reported. Figure~\ref{fig:cleaning} orients
the reader to the Knott's layout of pathways, commercial zones, and attraction envelopes, and contrasts
the full set of raw imported trajectories with the cleaned, pathway-only tracks that result.

For each visit $i$ we computed eight features, defined in Table~\ref{tab:features}. The notation is as
follows: $P_i$ is the set of POI visits, $A_i\subseteq P_i$ the attractions, $T_i$ the total in-park
time, $t^{\text{path}}_i$ the time spent on pathways, $S_i$ the pathway GPS segments with
importer-derived speeds $v_s$, $\Delta_a$ the dwell time inside attraction $a$, and
$d^{\text{walk}}_j$, $d^{\text{short}}_j$ the walked and shortest-path distances of inter-POI leg $j$,
of which there are $m$; $\tau^{\text{entry}}_i$ and $\tau^{\text{first}}_i$ are the times of park entry
and first attraction entry.

\begin{table}[t]
\centering
\caption{The eight movement features, their definitions, units, and the direction in which each is
hypothesized to shift from waders toward divers. Features are computed per visit; see text for notation.}
\label{tab:features}
\small
\begin{tabular}{p{0.3\linewidth} p{0.20\linewidth} l p{0.155\linewidth}}
\toprule
Feature & Definition & Units & Hypothesized diver tendency \\
\midrule
\% time on paths                              & $t^{\text{path}}_i / T_i$ & fraction & lower \\
Average speed                                 & $\tfrac{1}{|S_i|}\sum_{s\in S_i} v_s$ & m\,s$^{-1}$ & higher \\
POIs per hour                                 & $|P_i| / T_i$ & h$^{-1}$ & higher \\
Attractions per hour                          & $|A_i| / T_i$ & h$^{-1}$ & higher \\
Average attraction duration\textsuperscript{a} & $\tfrac{1}{|A_i|}\sum_{a\in A_i}\Delta_a$ & min & lower \\
Path directness\textsuperscript{b}           & $\tfrac{1}{m}\sum_{j} d^{\text{short}}_j / d^{\text{walk}}_j$ & ratio & higher \\
Attraction focus                              & $|A_i| / |P_i|$ & ratio & higher \\
Time to first attraction                      & $\tau^{\text{first}}_i - \tau^{\text{entry}}_i$ & min & lower \\
\bottomrule
\end{tabular}

\smallskip
{\footnotesize\raggedright \textsuperscript{a}Standardized to a within-day $z$-score at Knott's Berry
Farm to control for between-day crowd differences (see text); raw units are minutes.
\textsuperscript{b}Ratio of shortest-path distance to walked distance, so $1.0$ indicates a direct route
and smaller values more circuitous travel.\par}
\end{table}

The shortest-path distance for path directness was computed on a Delaunay-triangulated pathway network
buffered \SI{1}{\metre} inward, so that a value near 1.0 indicates a direct route and smaller values
indicate circuitous travel. Because mean attraction duration differed sharply between the two Knott's
days (about \SI{27}{\minute} versus \SI{16.5}{\minute}, attributable to differing crowd levels),
average attraction duration was converted to a within-day $z$-score for Knott's only, so that it reflects
relative rather than absolute waiting.

The eight features were chosen so that, taken together, they triangulate the diving-to-wading gradient
from several behavioral angles, rather than relying on any single proxy. Three capture intensity of
engagement: POIs per hour and attractions per hour should be higher for divers, who try to see as much
as possible, and the attraction-focus ratio should be higher as well, because diver visits are
weighted toward rides rather than incidental retail. Two capture pace and decisiveness: average speed
and path directness should be higher for divers, who move purposefully between chosen destinations,
whereas waders meander and backtrack. Two capture the temporal envelope of the visit: time to first
attraction should be short for divers, who head straight for a ride, and average attraction duration
should be short for divers at a park where dwelling trades against ride count, because divers tolerate
shorter experiences in exchange for more of them. The eighth feature, percentage of time on paths, was
included as a global summary of how much of the day is spent in transit rather than inside envelopes,
with an expectation, which the data later overturn, that divers spend less of their day on paths. We
report all eight despite their partial redundancy, both for transparency and because the redundancy is
itself informative about the latent structure of the behavior, but we treat the rate and focus features
as the primary carriers of the diving signal and interpret the others in their light.

\subsection{Clustering and multi-criteria validation}
Features were standardized to zero mean and unit variance so that no feature dominated by scale, then
clustered with $k$-means (scikit-learn, \texttt{random\_state}=42, \texttt{n\_init}=10), separately for
each park. We chose $k=3$ \emph{a priori} to match the industry typology; this is an interpretive choice
rather than a data-optimal one, and we report validity across $k=2$ to $5$ to make the trade-off
explicit. The gap statistic was computed to probe this assumption, not to select $k$; as the Results show,
it identifies no natural number of clusters, which is consistent with treating $k=3$ as a practical
discretization of a continuum rather than as a recovered optimum. Following the framework of Section~\ref{sec:framework}, each solution was assessed on three
distinct properties. \emph{Separation} was measured by the silhouette coefficient and the Davies--Bouldin
and Caliński--Harabasz indices.
\emph{Reproducibility} was measured in two complementary ways: restart stability, defined as the mean
adjusted Rand index (ARI) between the canonical solution and 300 random-initialization runs; and
subsampling consensus \parencite{Monti2003Consensus}, in which $k$-means was applied to 1{,}000 random
subsamples of 80\% of participants and the fraction of subsamples in which each pair of participants was co-clustered was
recorded as a consensus matrix, summarized by the mean within-cluster and between-cluster co-assignment.
\emph{Algorithm robustness} was measured by the ARI between the $k$-means partition and agglomerative
(Ward linkage) and Gaussian-mixture solutions on the same features. We specified, before inspecting
cluster contents, a single coherence rule (silhouette $\ge 0.35$ \emph{and} mean restart ARI $\ge 0.60$
at $k=3$), so that the decision to report a park quantitatively did not depend on whether its clusters
happened to look interpretable. The silhouette threshold of 0.35 follows the conventional reading of
\textcite{Rousseeuw1987Silhouettes}, on which values below about 0.25 indicate negligible structure and values in
the 0.25 to 0.50 range indicate weak-to-moderate structure; we placed the bar in the middle of that
band, acknowledging that behavioral continua rarely yield sharply separated clusters. We also examined the Spearman correlation structure among the eight
features and the variance explained by the leading principal components, because multicollinearity
bears on how to interpret which features drive the partition.

Labeling clusters as waders, swimmers, or divers risks circularity if the same quantity drives both the
assignment and its validation. We therefore separated the two. We first stated the expected
directionality of each feature \emph{a priori} (for example, divers should be faster, more
attraction-focused, and quicker to reach a first attraction). We then ran $k$-means blind to labels,
ordered the resulting clusters by their mean rank-average (the mean across features of each
participant's percentile rank, sign-flipped for diver-low features), and assigned wader, swimmer, and
diver in ascending order. Only after labeling did we test, independently, whether the groups actually
differ on the raw features (Kruskal--Wallis, with $\eta^2$ effect size) and whether they align with the
exit survey. The rank-average only orders the clusters; the Kruskal--Wallis tests and the survey provide
validation that does not reuse the labeling logic. Bootstrap 95\% confidence intervals ($10^4$ resamples)
summarize uncertainty in cluster feature means.

\subsection{Survey validation and cross-park directionality}
To test external validity we joined four numeric survey variables (total planning score, self-described
movement, months since last visit, and map-consultation frequency) to the $k$-means cluster labels by
participant identifier, so that the survey played no part in forming the clusters, and we tested
association with both Kruskal--Wallis across the three groups and a Spearman correlation between the
ordinal cluster label and each variable. To quantify cross-park directionality we computed, within each
park, the standardized (diver minus wader) difference of each feature in standard-deviation units and
recorded its sign, then counted the features whose sign reverses between Knott's and each Disney park.

\subsection{Agent-based demonstration}
Finally we asked whether the typology carries information useful for simulation, and in particular
whether the difference between visitor types lives in \emph{how} they walk or in \emph{where} they
choose to go. We built a destination-choice model on the real Knott's pathway network, a graph of
$20{,}236$ nodes derived from the park's path polygons. The 37 attractions were tagged with the park's
thrill-level layers (level 1, low, to level 5, aggressive), and commercial destinations were sampled from
the retail-and-dining footprint. Each agent makes a sequence of stops, their number set by the measured
POI rate; at each stop it chooses an
attraction (with the measured attraction-focus probability) or a commercial destination, and within that
category selects a specific destination $d$ with probability $P(d)\propto w_\theta(d)\,
\mathrm{dist}(d)^{-\gamma_\theta}$, where $\mathrm{dist}$ is network distance from the current location,
$\gamma_\theta$ is a type-specific distance discount, and $w_\theta$ is an a priori, type-specific
preference weight. The weights encode the typology as hypotheses rather than fitted quantities: divers
tilt toward high-thrill rides, waders toward low-thrill and commercial destinations, and swimmers toward
moderate rides, each with a modest (about three-to-one) odds ratio across the thrill range. Visit
sequencing is captured by a zone-persistence term that makes waders chain stops within the same park zone
and divers roam between zones. Because park time is dominated by queuing, riding, and browsing rather than
walking, each visit deposits a dwell occupancy at its destination in addition to the transit time
accumulated along the route; fields are rasterized with the same kernel-density settings as the observed
heatmaps. The locomotion layer follows the social-force tradition \parencite{Helbing1995Social}.

We compared three agent populations under identical network and rasterization settings: a
\emph{locomotion-only} model whose agents carry the measured per-type parameters (speed, distance
discount, visit rate, attraction focus) but no destination preference or sequencing; the \emph{full}
model above; and a \emph{homogeneous} baseline whose agents all share the mean parameters. Rather than correlate full heatmaps pixel by pixel,
a comparison that heatmaps built from three to seven participants cannot support, we evaluated an
interpretable summary statistic: the thrill-versus-commercial occupancy gradient, defined for each type as its relative
occupancy near high-thrill rides minus its relative occupancy in commercial areas. We report this
gradient for the observed heatmaps and for each model; the diver-minus-wader separation, with
\SI{95}{\percent} intervals over 16 random seeds; and, in the Supplementary Material, the weaker
pixelwise comparison. The preference
weights are stated a priori from the typology and are not fit to the observed heatmaps, so the
demonstration is a forward test of a mechanism, not a calibrated reconstruction.

\paragraph{Reproducibility.} All analyses are fully scripted, and every statistic and figure reported in
Section~\ref{sec:results}, together with the agent-based simulation, can be regenerated from the
archived materials (see the Data Availability statement). Released GPS data are de-identified, recording
local time-of-day without calendar dates and identifying participants only by code.

\section{Results}\label{sec:results}

\subsection{Separation is weak, but the Knott's partition is reproducible}
\begin{figure}[t]
\centering
\includegraphics{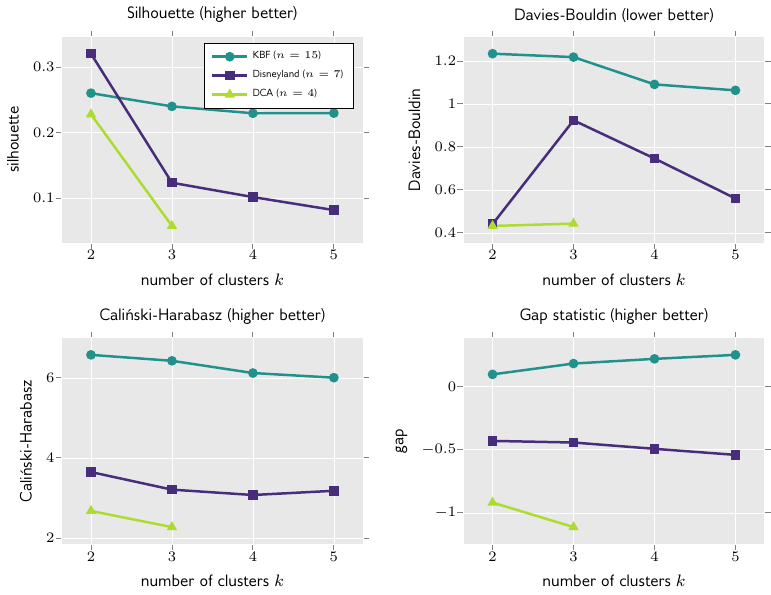}
\caption{Cluster-validity metrics versus $k$ for each park. Knott's is the least poorly separated yet
still falls below the silhouette coherence threshold (dashed line); the Disney samples show no clear
structure.}
\label{fig:validity}
\end{figure}

No park yields strongly separated clusters (Table~\ref{tab:validity}; Figure~\ref{fig:validity}). At
Knott's the $k=3$ silhouette is only 0.24 (Davies--Bouldin 1.22, Caliński--Harabasz 6.4), and the gap
statistic increases monotonically with $k$ rather than peaking, indicating no single natural number of
clusters. The Disney samples are weaker still at $k=3$ (silhouette 0.12 at Disneyland, 0.06 at DCA), and
their gap values are negative, that is, worse than a uniform null. By the a priori rule, no park is
``coherent'' on the separation criterion at $k=3$.

\begin{figure}[t]
\centering
\includegraphics{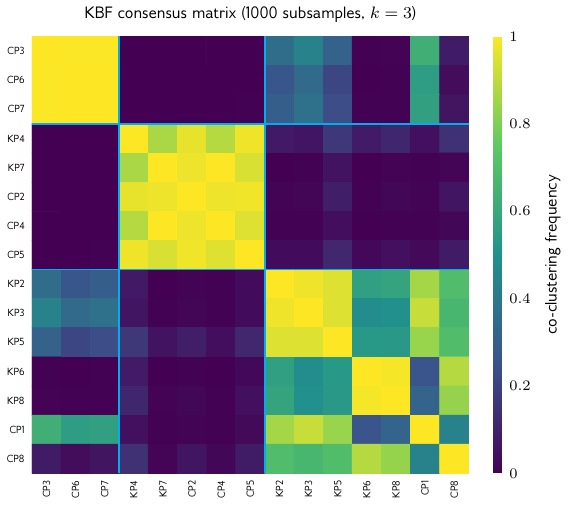}
\caption{Knott's subsampling consensus matrix (1{,}000 subsamples of 80\% of participants, $k=3$). Cells give the frequency
with which each participant pair is co-clustered; cyan lines mark the canonical clusters. Within-cluster
co-assignment averages 0.79 versus 0.09 between clusters, indicating a reproducible partition despite
weak geometric separation.}
\label{fig:consensus}
\end{figure}

\begin{figure}[t]
\centering
\includegraphics{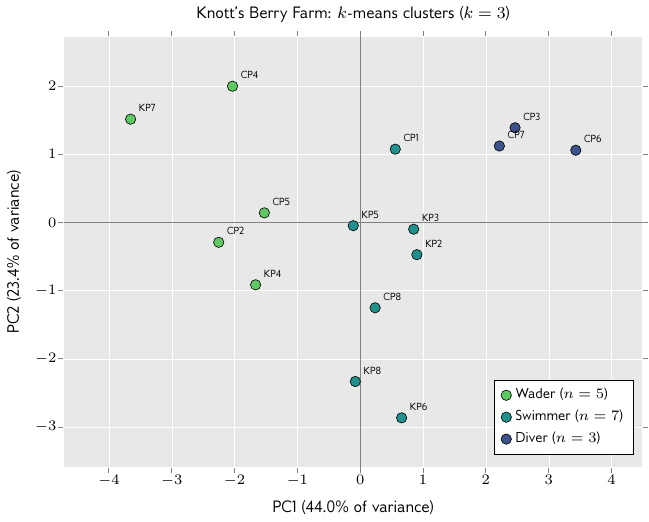}
\caption{Knott's $k=3$ clusters in the space of the first two principal components (labeled by
rank-average). Participants array along PC1, which carries 44.0\% of variance, with waders, swimmers,
and divers occupying successive regions and no clean gap between them, consistent with a behavioral
continuum partitioned into bands rather than discrete clusters.}
\label{fig:pca}
\end{figure}

Reproducibility, however, tells a sharply different story at Knott's, and this dissociation is one of
our central methodological findings. Restart stability is moderate (mean $\pm$ SD ARI over 300 restarts: $0.62\pm0.28$ at Knott's,
$0.36\pm0.25$ at Disneyland, and a spuriously high $0.89\pm0.35$ at DCA, the last an artifact of $n=4$
admitting essentially one partition). The stronger evidence comes from two further tests. First,
agglomerative (Ward) clustering returns \emph{exactly} the same three Knott's groups as $k$-means (ARI
$=1.00$); a Gaussian-mixture solution, which models within-cluster covariance, agrees only moderately
(ARI $=0.48$). Exact agreement between two algorithms with different inductive biases is hard to
attribute to a method artifact. Second, subsampling consensus over the 1{,}000 subsamples
(Figure~\ref{fig:consensus}) shows a clean block structure: participants assigned to the same cluster are
co-clustered in 0.79 of subsamples on average, versus 0.09 for participants in different clusters, a
nine-to-one ratio. The diver group in particular is recovered in essentially every subsample. Taken
together, the Knott's partition is highly reproducible even though it is not geometrically well
separated. Principal-component analysis clarifies why (Figure~\ref{fig:pca}): the first component alone
explains 44.0\% of feature variance and the second a further 23.4\%, for a cumulative 67.4\%, so the participants array along a
dominant behavioral gradient rather than into isolated clouds. The projection makes the structure
visible: waders, swimmers, and divers occupy successive regions along the first component with no clean
gap between them, the spatial expression of bands drawn across a continuum. Weak separation with high
reproducibility is the quantitative signature of such a continuum partitioned into convenient bands,
which echoes \textcite{Beeco2013GPS}, who found a binary wanderer/planner split insufficient, and is
exactly the situation our framework anticipated (E1, E3).

\begin{table}[t]
\centering
\caption{Cluster-validity indices by park (canonical $k$-means). Higher silhouette,
Caliński--Harabasz, and gap, and lower Davies--Bouldin, indicate better-separated clusters. No park
reaches the a priori silhouette threshold of 0.35 at $k=3$.}
\label{tab:validity}
\small
\begin{tabular}{llcccc}
\toprule
Park & $k$ & Silhouette & Davies--Bouldin & Caliński--Harabasz & Gap \\
\midrule
Knott's ($n{=}15$)    & 2 & 0.26 & 1.23 & 6.57 & 0.10 \\
                      & 3 & 0.24 & 1.22 & 6.42 & 0.18 \\
Disneyland ($n{=}7$)  & 2 & 0.32 & 0.44 & 3.65 & $-0.43$ \\
                      & 3 & 0.12 & 0.92 & 3.21 & $-0.44$ \\
DCA ($n{=}4$)         & 2 & 0.23 & 0.43 & 2.68 & $-0.92$ \\
                      & 3 & 0.06 & 0.44 & 2.27 & $-1.11$ \\
\bottomrule
\end{tabular}
\end{table}

\subsection{Feature correlation structure}
\begin{figure}[t]
\centering
\includegraphics{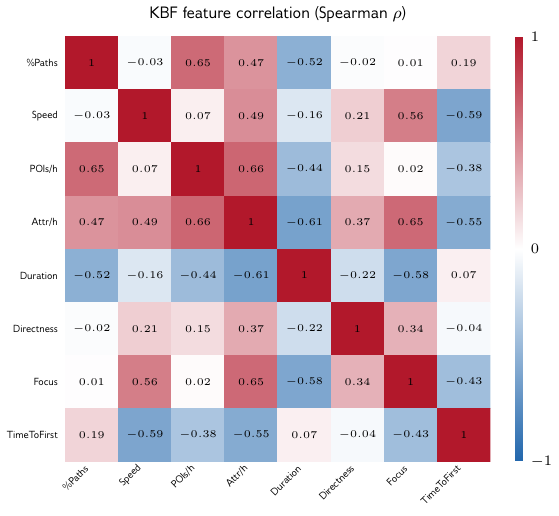}
\caption{Spearman correlation among the eight movement features at Knott's. The visitation-rate and
attraction-focus features form a correlated block, while directness, speed, and time-to-first-attraction
are comparatively independent.}
\label{fig:corr}
\end{figure}

The eight features are not independent (Figure~\ref{fig:corr}). The diving signal sits in a correlated
rate-block: POIs per hour correlates with attractions per hour (Spearman $\rho=0.66$) and with percentage
of time on paths ($\rho=0.65$), attractions per hour correlates with attraction focus ($\rho=0.65$) and
negatively with attraction duration ($\rho=-0.61$), while directness, speed, and time to first
attraction are comparatively independent. This multicollinearity means that the partition is driven less
by eight separate dimensions than by a smaller number of latent contrasts, consistent with the
dominance of the first principal component, and it cautions against interpreting each feature as
independent evidence. It also implies that a more parsimonious feature set would likely recover the same
partition, a useful simplification for future, larger studies.

\subsection{Knott's groups differ on behavioral rates}
Modest geometric separation does not mean the groups are interchangeable. At Knott's the $k=3$ partition
(divers $n=3$, swimmers $n=7$, waders $n=5$) produces groups that differ on the rate-based features that
most directly encode the typology (Table~\ref{tab:kbf}). By Kruskal--Wallis across the three groups, the
smallest uncorrected $p$-values are for attractions per hour ($H=10.9$, $p=0.004$, $\eta^2=0.74$),
attraction-focus ratio ($H=9.1$, $p=0.011$), walking speed ($H=8.0$, $p=0.019$), POIs per hour
($p=0.034$), and attraction duration ($p=0.035$). Because eight features are tested, we apply a
Benjamini--Hochberg false-discovery-rate correction: two features survive at the 0.05 level, attractions
per hour ($p_{\mathrm{BH}}=0.035$) and attraction focus ($p_{\mathrm{BH}}=0.043$), while walking speed,
POIs per hour, and attraction duration become marginal after correction ($p_{\mathrm{BH}}$ between 0.05
and 0.06). At this sample size we treat all of these as exploratory and read the corrected result as
identifying attraction-visitation rate and attraction focus as the most robust discriminators. The
pattern is nonetheless in the expected direction throughout: divers visit nearly five times as many
attractions per hour as waders (2.9 versus 0.6) and are far more attraction-focused, while waders dwell
longer in the attractions they do enter and reach their first attraction much later (means of 64 versus
11 minutes, a difference that is not significant, $p=0.21$). Path directness does not separate the groups
($p=0.33$), and percentage of time on paths is only marginal ($p=0.07$) and, as discussed in
Section~\ref{sec:disc}, runs opposite to the naive expectation. These differences on theoretically
motivated features, even though only two survive correction at this sample size, are in the direction
expectation E2 anticipated and provide the behavioral content that the weak silhouette alone would not
reveal.

\begin{table}[t]
\centering
\caption{Knott's $k=3$ cluster feature profiles: mean [bootstrap 95\% CI], raw Kruskal--Wallis $p$, and
Benjamini--Hochberg FDR-adjusted $p_{\mathrm{BH}}$ across the three groups. Attraction duration is a
within-day $z$-score (see Methods); other units as noted. Bold marks the two features that survive FDR
correction at $p_{\mathrm{BH}}<0.05$.}
\label{tab:kbf}
\small
\begin{tabular}{lccccc}
\toprule
Feature & Wader ($n{=}5$) & Swimmer ($n{=}7$) & Diver ($n{=}3$) & KW $p$ & $p_{\mathrm{BH}}$ \\
\midrule
Attractions / h        & 0.60 [0.45,\,0.76] & 1.46 [1.21,\,1.63] & 2.93 [2.49,\,3.72] & 0.004 & \textbf{0.035} \\
Attraction-focus ratio & 0.35 [0.23,\,0.50] & 0.77 [0.68,\,0.86] & 0.78 [0.63,\,1.00] & 0.011 & \textbf{0.043} \\
Avg.\ speed (m/s)      & 0.53 [0.51,\,0.56] & 0.63 [0.59,\,0.68] & 0.64 [0.61,\,0.68] & 0.019 & 0.050 \\
POIs / h               & 1.97 [1.32,\,2.62] & 1.94 [1.57,\,2.28] & 3.80 [3.56,\,4.13] & 0.034 & 0.057 \\
Avg.\ attraction dur.\ ($z$) & 0.96 [0.58,\,1.38] & $-0.30$ [$-0.80$,\,0.26] & $-0.91$ [$-1.10$,\,$-0.59$] & 0.035 & 0.057 \\
\% time on paths       & 0.26 [0.22,\,0.30] & 0.24 [0.18,\,0.30] & 0.36 [0.32,\,0.41] & 0.068 & 0.090 \\
Time to first attr.\ (min) & 64 [24,\,110] & 15 [7,\,25] & 11 [6,\,20] & 0.209 & 0.239 \\
Path directness        & 0.66 [0.61,\,0.70] & 0.68 [0.62,\,0.75] & 0.75 [0.68,\,0.85] & 0.329 & 0.329 \\
\bottomrule
\end{tabular}
\end{table}

\subsection{The exit survey does not validate the clusters}
Expectation E4, that self-report would corroborate the movement clusters, is \emph{not} supported, and
we report this null straightforwardly because it is informative. Joining the four numeric survey
variables to the $k$-means labels at Knott's, none differs significantly across the three groups
(Table~\ref{tab:survey}): planning score (Kruskal--Wallis $p=0.33$; Spearman $\rho=0.40$, $p=0.14$),
self-described movement ($p=0.79$; $\rho=0.11$), months since last visit ($p=0.44$; $\rho=0.19$), and
map use ($p=0.44$; $\rho=0.03$). Several group means are not even monotonic in the expected direction:
the GPS-defined waders report slightly \emph{higher} planning than the GPS-defined divers (1.33 versus
1.00), and self-description barely varies across groups. The most plausible reading is that objective
movement style and stated planning or self-image diverge, a divergence consistent with the warning in
the movement literature that travelers describe themselves imperfectly and shift style within a visit
\parencite{Mckercher2008Movement,Beeco2013GPS}. It also reflects limited statistical power at $n=15$. The practical
implication is cautionary: pre-visit survey items of the kind tested here are not a reliable proxy for
the movement types recovered from GPS, so segmentation for simulation should be grounded in movement data
rather than in self-report.

\begin{table}[t]
\centering
\caption{Quantitative external validation at Knott's: exit-survey variables by $k$-means cluster (survey
not used in clustering). No variable differs significantly across clusters; several are non-monotonic.}
\label{tab:survey}
\small
\begin{tabular}{lccccc}
\toprule
Survey variable & Wader & Swimmer & Diver & KW $p$ & Spearman $\rho$ ($p$) \\
\midrule
Planning score (0--2)        & 1.33 & 0.70 & 1.00 & 0.33 & 0.40 (0.14) \\
Self-description (0--1)      & 0.33 & 0.20 & 0.36 & 0.79 & 0.11 (0.71) \\
Months since last visit      & 4.34 & 0.70 & 4.47 & 0.44 & 0.19 (0.52) \\
Map-consultation (0--1)      & 0.33 & 0.40 & 0.71 & 0.44 & 0.03 (0.91) \\
\bottomrule
\end{tabular}
\end{table}

\subsection{Spatial patterns at Knott's}
\begin{figure}[t]
\centering
\includegraphics[width=\linewidth]{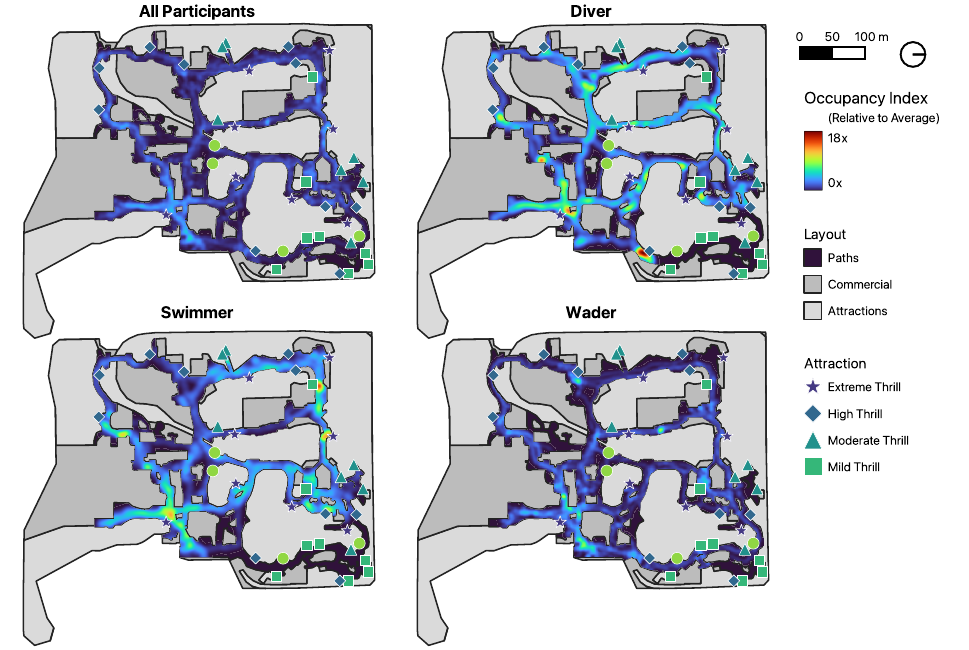}
\caption{Knott's occupancy heatmaps (relative to study average) for all participants and for each
cluster. Diver density favors thrill-attraction nodes; wader density favors commercial zones.}
\label{fig:heatmap}
\end{figure}

Per-cluster kernel-density heatmaps (Figure~\ref{fig:heatmap}), weighted by the inverse of each cluster's
total pathway time and scaled to multiples of the park-average occupancy of $3.42~\occunit$, are consistent with the
feature profiles. Diver density concentrates at the entrances and exits of high-thrill attractions,
wader density in commercial zones such as Ghost Town and Fiesta Village, and swimmer density falls
between the two. We report heatmaps descriptively only: spatial autocorrelation precludes formal
pixel-level testing, and one diver hotspot is a known single-participant artifact of the small sample, a
reminder that at this scale individual visitors can dominate the aggregate.

\subsection{Cross-park directionality reverses: the central finding}\label{sec:cross}
\begin{figure}[t]
\centering
\includegraphics{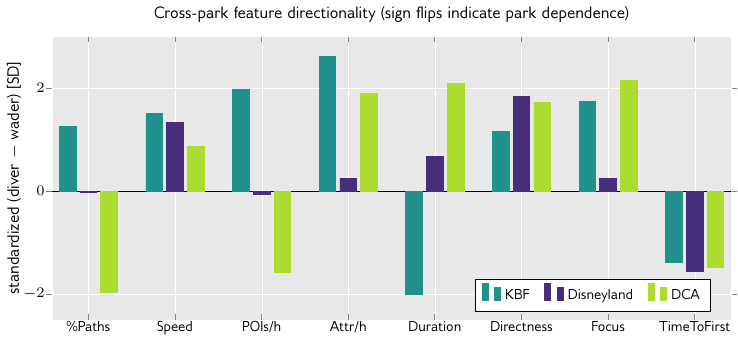}
\caption{Standardized (diver minus wader) difference per feature per park, in standard-deviation units.
The Knott's ($n=15$) partition is reproducible across algorithms and resamples, whereas the Disneyland
($n=7$) and DCA ($n=4$) partitions are not and are shown for exploratory comparison only. Five features keep the same sign
across parks (robust discriminators); three reverse between Knott's and the Disney parks
(park-contingent), most strikingly average attraction duration.}
\label{fig:direction}
\end{figure}

The headline result is that the \emph{sign} of cluster differences depends on the park
(Figure~\ref{fig:direction}). Of the eight features, five act as robust cross-park discriminators whose
standardized diver-minus-wader difference keeps the same sign at all three parks: walking speed,
attractions per hour, path directness, attraction focus, and time to first attraction all separate
divers from waders in the same direction everywhere. Three features reverse sign between Knott's and the
Disney parks. The most striking is average attraction duration, which is the extreme \emph{lowest} value
for divers at Knott's (standardized difference $-2.0$) but the \emph{highest} for divers at DCA
($+2.1$). Percentage of time on paths and POIs per hour likewise flip between Knott's and DCA. A
$k$-means run on the pooled three-park data is correspondingly incoherent: participants who are
unambiguous divers within their own park scatter across the pooled clusters (Supplementary Material).
This sharpens the park-dependence claim: a planner can reuse a subset of discriminators across
parks of broadly similar type, but cannot reuse the reversing features, and certainly cannot fit one
model to a mixture of park types. Expectation E5 is supported, and the result is the one with the
clearest practical consequence.

\subsection{Agent-based demonstration: destination choice, not locomotion, reproduces the type signature}\label{sec:abm}
\begin{figure}[t]
\centering
\includegraphics[width=\linewidth]{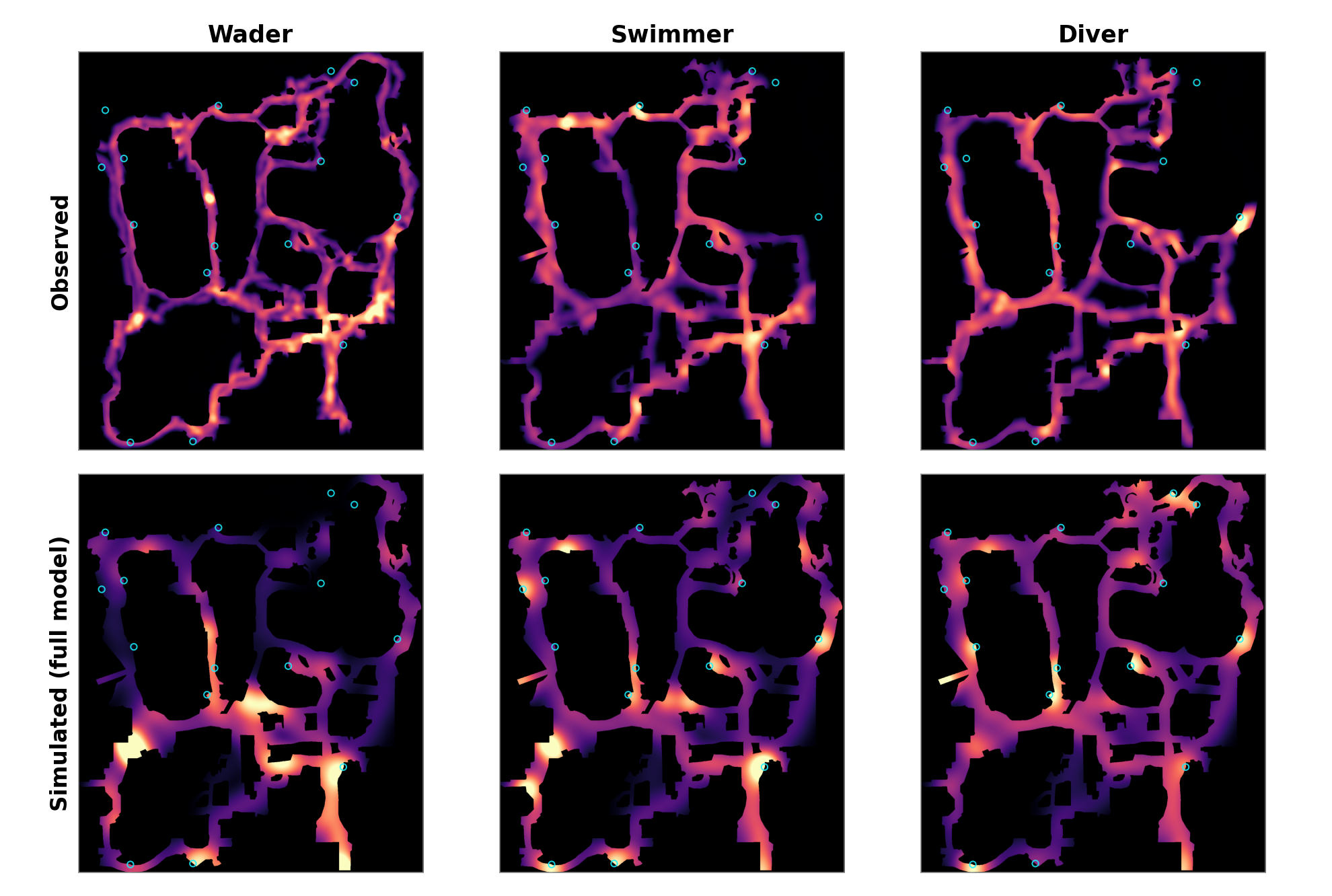}
\caption{Knott's pathway occupancy by visitor type, relative to the park average, for the observed
cluster heatmaps (top) and the full-model simulation (bottom), on a shared color scale; brighter is higher
occupancy and cyan rings mark high-thrill (level 4--5) rides. The simulated field is built like the
observed one, as continuous pathway presence (walking plus queue time laid along the approach), with
agents spread across near-optimal route variants; its type-specific destination preferences are set a
priori and are not fit to these heatmaps. Both rows show movement filling the path network, and each
type's occupancy concentrates where the observed heatmaps do, with divers around the high-thrill rides and
waders toward the commercial cores. The quantitative type signature and the three-model comparison are
reported in the Supplementary Material (\S2).}
\label{fig:abm-maps}
\end{figure}

The applied test (RQ4) yields a sharper result than a binary success would. We first establish,
model-free, that the visitor types leave a spatial signature beyond their aggregate features. The
observed Knott's heatmaps show a monotone thrill-versus-commercial occupancy gradient
(top row, Figure~\ref{fig:abm-maps}): relative to the park average, waders occupy high-thrill ride areas at $0.96$
and commercial areas at $0.35$ (gradient $+0.61$), swimmers at $1.28$ and $0.46$ (gradient $+0.82$), and
divers at $1.43$ and $0.24$ (gradient $+1.19$). Divers concentrate where the thrill is and avoid retail;
waders do the reverse; swimmers fall between. The diver-minus-wader separation is $+0.58$. This gradient
is measured directly from the data and does not depend on any model.

We then ask which modeling ingredients reproduce it. The discriminating statistic is the
diver-minus-wader separation with its \SI{95}{\percent} interval over 16 random seeds. A locomotion-only
population, whose agents carry the measured speed, distance discount, visit rate, and attraction focus
but no destination preference, does not reproduce the gradient: its separation is only $+0.14$ (interval $-0.23$ to $+0.42$, including zero), and its
per-type profile peaks at swimmers rather than rising to divers. A homogeneous baseline is flatter still
($+0.06$, $-0.32$ to $+0.39$). The full model, which adds a priori type-specific destination preference
and visit sequencing, is the only one whose separation excludes zero ($+0.99$, $+0.58$ to $+1.30$) and
whose profile rises monotonically from waders to divers as the observed one does. The heterogeneity that
matters, in other words, is not in how the types walk or how much they visit but in where they choose to
go: the locomotion-only population already carries every measured difference in Table~\ref{tab:abm}, yet
without destination preference those differences fail to separate the types along the
thrill--commercial axis. In
the maps, the full model's simulated occupancy reproduces the observed pattern, glowing along the same
corridors and concentrating each type's dwell where the observed heatmaps do, with divers clustered at the
high-thrill rides and waders spread toward the commercial cores (bottom row, Figure~\ref{fig:abm-maps}).

Two qualifications attach to the positive result. First, the full model somewhat overstates the magnitude
of the gradient (separation $+0.99$ versus the observed $+0.58$): the preference weights are set a priori
from the typology, not calibrated to the heatmaps, so the demonstration shows that type-specific
destination choice is sufficient and, within these three models, necessary to produce the observed
signature, not that these particular weights are correct. Second, reproducing the full per-cluster heatmaps pixel by
pixel remains beyond what $n=3$--$7$-participant heatmaps support; all three models have near-zero
pixelwise diagonal correlation, which is why we evaluate the gradient the data can resolve rather than the
pixels it cannot. With those caveats, Expectation E6 is supported in qualified form: a typed agent population improves on
a homogeneous baseline, but only once the types differ in destination preference and visit sequencing,
not in measured movement parameters alone. The full per-model gradients, the three-model comparison, and
the pixelwise check are tabulated and plotted in the Supplementary Material (\S2).

\begin{table}[t]
\centering
\caption{Agent parameters by visitor type. Locomotion and visit-budget parameters are the measured
Knott's cluster means; the destination-choice and sequencing parameters are a priori
hypotheses from the typology (divers tilt toward high-thrill rides, waders toward low-thrill and
commercial destinations, swimmers toward moderate rides; waders chain stops within a park zone, divers
roam). Each visit also lays its queue/dwell time ($\approx$4~min at attractions, 2~min at commercial
stops) along the approaching pathway, and agents draw among eight near-optimal route variants so that
movement spreads across the path network. These are hypotheses for simulation experiments, not validated
operational settings.}
\label{tab:abm}
\small
\begin{tabular}{lccc}
\toprule
Parameter (source) & Wader & Swimmer & Diver \\
\midrule
\multicolumn{4}{l}{\emph{Locomotion and visit budget (measured cluster means)}}\\
\quad Desired speed $v^0$ [m/s] (avg.\ speed)             & 0.53 & 0.63 & 0.64 \\
\quad Attraction-focus probability (attr.\ focus)         & 0.35 & 0.77 & 0.78 \\
\quad Visit rate [POI/h] (POIs per hour)                  & 1.97 & 1.94 & 3.80 \\
\quad Distance discount $\gamma_\theta$ (path directness) & 3    & 2    & 1    \\
\midrule
\multicolumn{4}{l}{\emph{Destination choice and sequencing (a priori)}}\\
\quad Thrill preference $w_\theta$ (level $1\!\to\!5$)    & $3\!\to\!1$ & peak at 3 & $1\!\to\!3$ \\
\quad Zone persistence (visit sequencing)                 & 3.0  & 1.7  & 1.0  \\
\bottomrule
\end{tabular}
\end{table}

\section{Discussion}\label{sec:disc}

\subsection{In what sense the typology exists}
Our results let us answer the question raised in Section~\ref{sec:framework} with more precision than the
existing literature. At the best-sampled park the wader/swimmer/diver typology is \emph{recoverable} and
\emph{reproducible}: the same three groups emerge from $k$-means and from Ward clustering identically,
they are recovered in roughly four of five subsamples, and they differ significantly on the movement
rates that define the types. At the same time the typology shows weak \emph{separation}: there is no gap
between the groups, and a single principal component carries almost half the variance. The honest
synthesis is that theme-park visitors occupy a continuous behavioral gradient that can be partitioned
into reproducible bands for practical purposes, not a set of naturally isolated clusters. This matters
because the segmentation and crowd-modeling literatures often report only one validity number, typically
a silhouette or an elbow, and would have dismissed this structure as ``no clusters.'' Distinguishing
reproducibility from separation recovers a usable, defensible segmentation from data that a single index
would discard, and it aligns the quantitative picture with the long-standing qualitative observation
that visitor types grade into one another \parencite{Beeco2013GPS,Mckercher2008Movement}.

\subsection{A transferable validation protocol for behavioral segmentation}
The decomposition of cluster quality into recoverability, reproducibility, separation, external
validity, and utility is, we argue, a contribution that travels beyond theme parks. Visitor and tourist
segmentation is a large and active field, yet a substantial share of published segmentations rest on a
single internal index, an elbow plot, or no stability assessment at all, and then proceed to interpret
and act on the resulting groups as if they were established entities. Our results show concretely why
that is hazardous and how cheaply it can be remedied. The Knott's partition would have been rejected by
a silhouette-only standard, yet it is reproducible across algorithms and resamples and behaviorally
meaningful; conversely, a partition can be reproducible while failing external validity, as the survey
result here demonstrates. Reporting these properties separately, using tools that are inexpensive and
already standard in adjacent fields, namely consensus clustering by subsampling \parencite{Monti2003Consensus},
cross-algorithm agreement via the adjusted Rand index, and independent-variable validation, lets an
analyst state precisely what kind of object a segmentation is before building on it. We would encourage
visitor-segmentation studies generally to report at least one separation diagnostic, one reproducibility
diagnostic, and one external-validity test, because the three answer different questions and, as we have
seen, can disagree. The protocol is also diagnostic in a constructive sense: the specific pattern of
weak separation with high reproducibility is not a null result but positive evidence for an underlying
continuum, a conclusion no single index could have delivered.

\subsection{A continuum, validated by movement rather than self-report}
The dissociation we observe speaks to an old tension in the field. Typological traditions from
\textcite{Cohen1972Sociology} and \textcite{Plog1974Why} through museum-visitor identity work \parencite{Falk2016Identity}
treat visitors as belonging to a small number of kinds, whereas movement-pattern research
\parencite{Lew2006Modeling,Mckercher2008Movement,Beeco2013GPS} repeatedly finds graded behavior and within-visit drift. Our
data suggest these views are reconcilable: theme-park movement varies continuously, but along a dominant
axis stable enough for practical bands to be drawn and reproduced. The industry
metaphor of waders, swimmers, and divers is, on this reading, neither simply right nor simply wrong; it
is a useful discretization of a gradient, with the important caveat that the cut-points between bands are
not natural breaks and may sit at different places at different parks. Treating the three types as crisp
categories with fixed boundaries invites over-confident planning, whereas treating them as labeled
regions of a continuum invites the more defensible practice of calibrating both the proportions and the
band locations to the specific park.

The continuum reading also makes sense of the survey null, which is a substantive finding rather than a
measurement failure. If visitors occupy a gradient rather than discrete kinds, coarse self-descriptions
such as wandering versus beelining will map onto it only loosely, and pre-visit planning, which concerns
intentions rather than realized movement, maps even more loosely. Pre-visit intentions and self-image
are, at the granularity and sample size studied here, weak predictors of how a visitor actually moves.
For research practice this argues that behavioral segmentation for pedestrian modeling should be built on
revealed movement rather than stated planning or self-description, with surveys better used to explain or
contextualize movement types than to define them; it also flags a validity caution for the many tourism
studies that segment visitors on questionnaire responses alone.

\subsection{Why directionality reverses}
The reversal of three features across parks is substantive and has a plausible mechanism. At a premium,
merchandise-rich resort, ``getting one's money's worth'' includes dwelling in shops, themed queues, and
immersive retail experiences, so a guest who is otherwise diver-like can spend \emph{longer} inside
attraction and commercial envelopes; at a regional park with lower per-capita spending, divers minimize
dwell in order to maximize ride count. Layout reinforces the effect: a hub-and-spoke topology that
funnels every guest through central commercial cores compresses the contrast between types, whereas a
more dispersed layout lets type differences express themselves. That speed, directness, attraction
focus, attractions per hour, and time to first attraction keep their sign across parks, while attraction
duration, percentage of time on paths, and POIs per hour reverse, suggests a practical division of the
feature set into transferable kinematic-and-intent measures and park-contingent dwell-and-exposure
measures. The core message for practice is that agent parameters must be calibrated on the specific type
of park being modeled rather than borrowed across venues, because the same measured feature can indicate
opposite behavior in different settings.

\subsection{An instructive feature failure}
One feature deserves separate comment because it overturned our prior in a revealing way. We had expected
divers to spend a smaller share of their day on pathways, on the reasoning that they waste less time in
transit; at Knott's the opposite held, with divers spending a larger share of the day on paths than
waders (Table~\ref{tab:kbf}). The mechanism is instructive. Waders, lacking a plan, tend to ride or shop
at whatever they encounter, chaining nearby points of interest over short distances, whereas divers,
having prioritized specific attractions, travel longer committed legs between them. Both aimless and
efficient behavior can therefore inflate pathway time, by different routes, so a single global
time-budget proxy of this kind cannot separate the types. This is why the rate and focus features, rather
than percentage of time on paths, carry the diving signal, and why we would exclude percentage of time on
paths from the parameter profile we recommend. The broader methodological lesson reinforces the previous
subsection: features must be interpreted in context, and an apparently intuitive proxy can encode the
opposite of what it seems to.

\subsection{What the simulation result implies}
The demonstration equips agents with the measured per-type profile (a visit budget from the rate and
focus features that carry the partition, Table~\ref{tab:kbf}, and locomotion from the speed and
directness features) and then asks what must be added before type-specific space use emerges. The answer
is clear and one-directional: the measured profile alone does not reproduce it, and a homogeneous
baseline does not, but adding a priori type-specific destination preference and visit sequencing does,
recovering the observed thrill-versus-commercial occupancy gradient in both direction and ordering
(Section~\ref{sec:abm}). This is consistent with the
correlation finding that the eight features collapse onto a dominant gradient, and with the observation
that all three observed heatmaps share the same pathway skeleton: the type signal lives not in the
skeleton or in the walking, but in which destinations each type dwells at. For simulation practice the
implication is concrete: the heterogeneity that matters for realistic space use lives in destination
choice and visit sequencing, the identity and order of the POIs a visitor selects, more than in how fast
or directly the visitor walks between them.

The model we use makes that claim operational. Rather than a fixed walking speed and goal, each agent
carries a type-specific destination-choice rule and a visit budget: a gravity-style rule weights
candidate destinations by a type-dependent combination of thrill attractiveness and network distance,
with divers placing high weight on high-thrill rides and low weight on distance, waders the reverse with
a stronger pull toward retail and low-thrill sights, and swimmers toward moderate rides; the visit budget
is set from the measured POIs-per-hour and attraction-focus profiles; and a zone-persistence process, a
type-specific transition over park zones in the spirit of the early flow models
\parencite{Ahmadi1997Managing,Rajaram2003Flow}, generates the ordered itinerary that the locomotion layer then
executes. Expressed this way, the model reproduces the heterogeneity where our results locate it, in what
is visited and in what order, while reusing the speed and directness parameters estimated here for the
movement between destinations. What remains is calibration and stronger validation: the a priori
preference weights recover the \emph{shape} of the observed gradient but overstate its \emph{magnitude},
and a full test of the type-specific structure would compare simulated and observed trajectories, for
instance by matching visit-sequence distributions or inter-destination transition matrices, rather than
the static occupancy gradient used here. Both steps require larger, better-powered samples than the
present pilot affords.

\subsection{Implications for practice}
For master planners and the vendors of pedestrian-simulation software, three concrete points follow.
First, a behavioral segmentation worth using need not be geometrically crisp; planners should judge a
proposed visitor typology by whether its groups are reproducible and differ on meaningful movement
features, not by a single separation index. Second, parameter libraries should be organized by park type
rather than offered as universal defaults, because the mapping from features to types is venue-specific;
a subset of discriminators may transfer among similar parks, but dwell-related features should be
recalibrated. Third, the pipeline itself, GPS plus a brief exit survey plus the eight features defined
here, is inexpensive enough to be run as a calibration step for a specific park before its movement model
is trusted for layout or safety decisions, and the analysis confirms that this calibration should be
grounded in movement data rather than survey self-report. This calibration step fills a specific gap:
the more flexible commercial pedestrian-modeling tools already let users set the proportion of waders,
swimmers, and divers in a park, and in limited ways how those types differ, but they offer no
empirically grounded values for those controls, leaving them to expert guesswork. The present pipeline
is a route to supplying such values for a given park type.

\subsection{Generalizability and boundary conditions}
Because the study is a small pilot, it is important to state where its conclusions should and should not
be expected to hold. The qualitative finding that movement is a reproducible continuum, and the
methodological argument for separating reproducibility from separation, are likely to generalize: they
follow from the structure of the data rather than from the particular parks, and the same dissociation
has analogues in other movement-segmentation settings. The specific Knott's parameter profiles, by
contrast, are tied to a regional, mixed-thrill park visited disproportionately by passholders in a
single season, and should be expected to shift for first-time visitors, for families with children, and
across seasons and crowding regimes; the passholder skew in particular likely inflates the wader-leaning
end of the gradient. The directionality-reversal finding is demonstrated here for one contrast, regional
versus premium-resort, and although the proposed mechanism, that premium pricing and immersive retail
turn dwelling into part of the diving experience, is general, the precise set of features that reverse
may differ for other park types such as water parks, indoor parks, or international resorts operating
under different reservation regimes. Reservation and virtual-queue systems are themselves a moving
target that reshapes flow \parencite{Pache2024Theme} and could alter which features discriminate types.
These boundary conditions are not caveats to be apologized for but a research agenda: each names a
comparison that a larger, multi-park, multi-season study could test directly, using exactly the pipeline
and validation protocol set out here.

\section{Limitations and future work}\label{sec:limits}
The sample is small ($n=15/7/4$), single-season, and non-representative. About half of participants had
visited within the prior month and were therefore likely passholders, and no families or children were
recruited; both biases skew the sample toward repeat, low-urgency, wader-leaning behavior, so the
profiles probably do not transfer to first-time or family visitors. Cluster separation is weak in
absolute terms, which is why we report $k=3$ as an interpretive choice, lean on reproducibility rather
than separation, avoid confirmatory hypothesis testing, and exclude the Disney parks from quantitative
recommendations. In particular, the diver group contains only three participants, so the Kruskal--Wallis
tests are underpowered and only two features survive multiple-comparison correction; the non-significant
features (path directness, time to first attraction) should be read as inconclusive rather than as
evidence of no difference, and larger samples are essential before any of these is treated as confirmed. Single-coder manual cleaning, the within-day $z$-scoring of one feature, the
multicollinearity among features, and the uncalibrated a priori weights of the agent-based demonstration
are further limitations. The survey null should be read as power-limited as well as substantive, and the
agent-based demonstration recovers the direction and ordering of the observed type gradient but, with
a priori weights, not its magnitude. Future
work should collect at least 50 participants per park across seasons and demographics, ideally through
virtual recruitment to reduce passholder and family bias; compute inter-rater reliability on track
cleaning; reduce the feature set in light of the observed correlation structure; extend the comparison to
more park types so that the directionality-reversal finding can be tested systematically; and, on the
modeling side, calibrate the destination-choice and sequencing sub-model demonstrated here against
held-out GPS trajectories, for instance by matching visit-sequence distributions, rather than relying on
the a priori weights and the static occupancy gradient used here.

\section{Conclusion}
The industry wader/swimmer/diver typology can be recovered from GPS traces, and at the best-sampled park
it is reproducible and behaviorally significant even though the groups are not geometrically sharp, which
indicates a continuum of movement styles rather than discrete types. Self-report does not distinguish the
groups, so segmentation for modeling should rest on revealed movement. Most consequentially, the
behavioral meaning of movement features is park-dependent: several features reverse direction between a
regional park and a premium resort, so visitor-behavior parameters cannot be universalized. An
agent-based test sharpens the point: measured movement parameters alone fail to separate the visitor
types on the observed thrill-versus-commercial occupancy gradient, whereas a model that adds
type-specific destination preference and visit sequencing reproduces it, locating the heterogeneity
worth modeling in destination choice rather than locomotion. This study contributes a reproducible, critically validated pipeline and
preliminary, park-specific evidence toward calibrating pedestrian simulations by park type; larger and
more representative data collection, together with destination-level modeling and simulation validation,
are the necessary next steps.


\section*{Disclosure statement}
The authors report no competing interests.

\section*{Use of generative AI}
During the preparation of this manuscript, the authors used Claude (Anthropic, Opus 4.8) to check spelling
and grammar and to correct typographical errors. The tool was not used to generate scientific content,
data, results, or interpretations. After using this tool, the authors reviewed and edited the text as
needed and take full responsibility for the content of the publication.

\section*{Data availability statement}
\ifdefined\anonymous
The de-identified data supporting this study, together with the analysis and agent-based-simulation code
that regenerates all data-driven figures, tables, and statistics in this article and its Supplementary
Material, are openly archived in a single public repository under open licenses (CC-BY-4.0 for the data,
MIT for the code). The repository citation is withheld here to preserve anonymity and will be provided
upon acceptance. GPS tracks are de-identified, recording local time-of-day without calendar dates and
identifying participants only by code. The recruitment, consent, and exit-survey instruments are
reproduced in the Supplementary Material.
\else
The de-identified data supporting this study are openly archived in a single Zenodo deposit
(DOI:~\href{https://doi.org/10.5281/zenodo.20748587}{10.5281/zenodo.20748587}), organized one folder per park: per-visit movement features, point-of-interest
visit timetables, exit-survey responses, GPS tracks (raw and cleaned), park GIS layers, and the observed
occupancy heatmaps. The same deposit includes, under \texttt{code/}, the analysis and
agent-based-simulation code that regenerates all data-driven figures, tables, and statistics in this
article and its Supplementary Material directly from the deposited data. The data are released under
CC-BY-4.0 and the code under the MIT License. GPS tracks are de-identified, recording local time-of-day
without calendar dates and identifying participants only by code. The recruitment, consent, and
exit-survey instruments are reproduced in the Supplementary Material.
\fi

\section*{Notes on contributors}
\ifdefined\anonymous
[Contributor details removed for anonymous peer review.]
\else
\emph{Dane M. Utley} conducted the fieldwork, data processing, analysis, and writing as a senior thesis
in Civil and Environmental Engineering at Princeton University. \emph{Jürgen Hackl} advised the study.
\fi

\printbibliography
\end{refsection}

\clearpage
\setcounter{section}{0}\renewcommand{\thesection}{S\arabic{section}}
\setcounter{figure}{0}\renewcommand{\thefigure}{S\arabic{figure}}
\setcounter{table}{0}\renewcommand{\thetable}{S\arabic{table}}
\setcounter{equation}{0}\renewcommand{\theequation}{S\arabic{equation}}

\begin{center}
  {\LARGE\bfseries Supplementary Information}\\[1.5ex]
  {\large Do Waders, Swimmers, and Divers Exist? A GPS-Based Pilot Study of Site-Dependent Visitor Movement in Theme Parks}
\end{center}
\bigskip

\begin{refsection}

This document provides (\ref{sec:methods}) the formal definitions and equations for every metric and
procedure used in the main text; (\ref{sec:support}) the full tables behind results the main text reports
in summary; and the material relegated from the main text for transparency, namely the two-cluster
comparison (\ref{sec:k2}), the under-powered Disneyland and Disney California Adventure (DCA) analyses
(\ref{sec:dldca}), the pooled cross-park model (\ref{sec:pooled}), the per-park rank-average tables
(\ref{sec:rank}), and the study instruments (\ref{sec:instruments}). None of the relegated material is
load-bearing for the main results. Notation follows the main text throughout.

\section{Supplementary methods: definitions and equations}\label{sec:methods}

\subsection{Movement features}
For a visit $i$, let $T_i$ be the total in-park time, $t^{\text{path}}_i$ the time on pathways, $P_i$ the
set of point-of-interest (POI) visits and $A_i\subseteq P_i$ the attractions, $S_i$ the pathway GPS
segments with importer-derived speeds $v_s$, $\Delta_a$ the dwell time inside attraction $a$, and, for
each of the $m_i$ inter-POI legs $j$, $d^{\text{walk}}_j$ and $d^{\text{short}}_j$ the walked and
shortest-path distances. The eight features are
\begin{align}
f_1 &= \frac{t^{\text{path}}_i}{T_i}, &
f_2 &= \frac{1}{|S_i|}\sum_{s\in S_i} v_s, &
f_3 &= \frac{|P_i|}{T_i}, &
f_4 &= \frac{|A_i|}{T_i}, \label{eq:feats1}\\[2pt]
f_5 &= \frac{1}{|A_i|}\sum_{a\in A_i}\Delta_a, &
f_6 &= \frac{1}{m_i}\sum_{j=1}^{m_i}\frac{d^{\text{short}}_j}{d^{\text{walk}}_j}, &
f_7 &= \frac{|A_i|}{|P_i|}, &
f_8 &= \tau^{\text{first}}_i-\tau^{\text{entry}}_i, \label{eq:feats2}
\end{align}
that is, percentage of time on paths ($f_1$), average speed ($f_2$), POIs per hour ($f_3$), attractions
per hour ($f_4$), average attraction duration ($f_5$), path directness ($f_6$), attraction focus ($f_7$),
and time to first attraction ($f_8$). The shortest-path distances in $f_6$ were computed on a planar path
graph obtained by Delaunay triangulation of tens of thousands of random points over each park's walkable
extent, clipped to the pathway polygon buffered \SI{1}{\metre} inward; $f_6=1$ denotes a perfectly direct
route. Because mean attraction duration differed between the two Knott's collection days, $f_5$ was
standardized within day for Knott's, $\tilde f_5 = (f_5-\mu_d)/\sigma_d$, with $\mu_d,\sigma_d$ the
day-$d$ mean and standard deviation.

\subsection{Standardization and clustering}
Each feature was standardized across participants, $z_{ij}=(x_{ij}-\bar{x}_j)/s_j$, so that no feature
dominates by scale. We partitioned the standardized vectors $\mathbf{z}_i$ with $k$-means
\parencite{MacQueen1967methods}, which minimizes the within-cluster sum of squares
\begin{equation}
\min_{C_1,\dots,C_k}\ \sum_{c=1}^{k}\sum_{i\in C_c}\bigl\lVert \mathbf{z}_i-\boldsymbol{\mu}_c\bigr\rVert^2,
\qquad \boldsymbol{\mu}_c=\frac{1}{|C_c|}\sum_{i\in C_c}\mathbf{z}_i ,
\label{eq:kmeans}
\end{equation}
using 10 random restarts and a fixed seed for the canonical solution.

\subsection{Internal validity indices}
For object $i$ let $a(i)$ be its mean distance to other members of its cluster and $b(i)$ the smallest
mean distance to any other cluster; the silhouette \parencite{Rousseeuw1987Silhouettes} is
\begin{equation}
s(i)=\frac{b(i)-a(i)}{\max\{a(i),b(i)\}}\in[-1,1],
\label{eq:sil}
\end{equation}
reported as the mean over objects. The Davies--Bouldin index \parencite{Davies1979Cluster} and the
Caliński--Harabasz index \parencite{Calinski1974dendrite} are
\begin{equation}
\mathrm{DB}=\frac{1}{k}\sum_{c=1}^{k}\max_{c'\neq c}\frac{\sigma_c+\sigma_{c'}}{d(\boldsymbol{\mu}_c,\boldsymbol{\mu}_{c'})},
\qquad
\mathrm{CH}=\frac{\operatorname{tr}(B_k)/(k-1)}{\operatorname{tr}(W_k)/(n-k)},
\label{eq:dbch}
\end{equation}
where $\sigma_c$ is the mean within-cluster distance to $\boldsymbol{\mu}_c$ and $B_k,W_k$ are the
between- and within-cluster scatter matrices. The number of clusters was probed with the gap statistic
\parencite{Tibshirani2001Estimating},
\begin{equation}
\mathrm{Gap}(k)=\mathbb{E}^{*}_{n}\!\left[\log W_k\right]-\log W_k,
\label{eq:gap}
\end{equation}
with the expectation taken over $B$ reference datasets drawn uniformly on the feature bounding box; the
gap was computed to probe, not to select, $k$.

\subsection{Reproducibility}
For two partitions with overlap counts $n_{ij}$ and row/column sums $a_i,b_j$, the adjusted Rand index is
\begin{equation}
\mathrm{ARI}=\frac{\sum_{ij}\binom{n_{ij}}{2}-\Bigl[\sum_i\binom{a_i}{2}\sum_j\binom{b_j}{2}\Bigr]\big/\binom{n}{2}}
{\tfrac{1}{2}\Bigl[\sum_i\binom{a_i}{2}+\sum_j\binom{b_j}{2}\Bigr]-\Bigl[\sum_i\binom{a_i}{2}\sum_j\binom{b_j}{2}\Bigr]\big/\binom{n}{2}}.
\label{eq:ari}
\end{equation}
We used the ARI in two ways: restart stability, the mean ARI between the canonical solution and 300
random-initialization runs; and cross-algorithm agreement, the ARI between $k$-means and agglomerative
(Ward) and Gaussian-mixture partitions. We also computed a subsampling consensus matrix
\parencite{Monti2003Consensus}: drawing $B=1000$ random $80\%$ subsamples and clustering each, the consensus is
\begin{equation}
C_{ij}=\frac{\#\{\text{runs in which $i,j$ are co-clustered}\}}{\#\{\text{runs in which $i,j$ are co-sampled}\}},
\label{eq:consensus}
\end{equation}
summarized by the mean within-cluster and between-cluster $C_{ij}$.

\subsection{Cluster labeling}
Labels were assigned without reusing the clustering. For feature $j$, let $r_{ij}\in[0,1]$ be participant
$i$'s percentile rank; for the three features on which a diver is expected to score \emph{low}
($f_1,f_5,f_8$) the rank is flipped, $\tilde{r}_{ij}=1-r_{ij}$. The rank-average is
\begin{equation}
R_i=\frac{1}{8}\sum_{j=1}^{8}\tilde{r}_{ij},
\label{eq:rankavg}
\end{equation}
and clusters are labeled wader, swimmer, diver in ascending order of their mean $R$.

\subsection{Inference}
Across the three clusters, each feature was compared with the Kruskal--Wallis test
\begin{equation}
H=\frac{12}{n(n+1)}\sum_{c=1}^{k}\frac{R_c^2}{n_c}-3(n+1),
\qquad \eta^2=\frac{H-k+1}{n-k},
\label{eq:kw}
\end{equation}
with $R_c$ the rank sum of cluster $c$ and $\eta^2$ the effect size. Because eight features are tested,
$p$-values were adjusted by the Benjamini--Hochberg false-discovery-rate procedure
\parencite{Benjamini1995Controlling}: ordering $p_{(1)}\le\cdots\le p_{(m)}$ with $m=8$, the adjusted values are
\begin{equation}
p^{\mathrm{BH}}_{(i)}=\min_{l\ge i}\min\!\left\{\frac{m}{l}\,p_{(l)},\,1\right\}.
\label{eq:bh}
\end{equation}
Uncertainty in cluster feature means is summarized by percentile bootstrap $95\%$ confidence intervals
from $10^4$ resamples.

\subsection{Cross-park directionality}
Within each park, features were standardized and the directionality of each feature was measured as the
standardized difference between the diver and wader cluster means,
\begin{equation}
\delta_j=\bar{z}^{\,\text{diver}}_j-\bar{z}^{\,\text{wader}}_j \quad[\text{SD units}].
\label{eq:direction}
\end{equation}
A change in the sign of $\delta_j$ between parks indicates that the behavioral meaning of feature $j$ is
park-contingent.

\subsection{Occupancy heatmaps}
For each cluster the cleaned pathway tracks were converted to vertex points; every point was weighted by
$1/\mathcal{T}$, where $\mathcal{T}$ is the cluster's total pathway time in hours, so that clusters of
different size and duration are comparable. A kernel-density estimate (radius \SI{7.5}{\metre}, pixel
\SI{0.25}{\metre}) was clipped to the pathways and rescaled to multiples of the park study-average pixel
value $\bar{\rho}$ (KBF $3.42$, Disneyland $3.74$, DCA $7.00~\occunit$), so that a value of $2$ means a
cluster is twice as likely as the sample average to occupy that pixel.

\subsection{Agent-based model}\label{sec:abm-methods}
Locomotion parameters were taken from the cluster means in the social-force tradition
\parencite{Helbing1995Social} (desired speed from $f_2$, attraction-focus probability from $f_7$, visit budget
from $f_3$ and $f_4$, distance discount $\gamma_\theta$ ordered by measured directness $f_6$). The 37
Knott's attractions were tagged with the park's thrill-level layers, giving each attraction a level
$\ell(d)\in\{1,\dots,5\}$ (low to aggressive); commercial destinations were sampled from the
retail-and-dining footprint. On the Knott's path graph ($20{,}236$ nodes), an agent of type $\theta$ at
location $u$ draws a category (attraction with probability $f_7$, else commercial) and selects a
destination $d$ within it with probability
\begin{equation}
P_\theta(d\mid u)\;\propto\; w_\theta\!\big(\ell(d)\big)\;\operatorname{dist}(u,d)^{-\gamma_\theta}\;
\pi_\theta^{\,\mathbb{1}[z(d)=z(u)]},
\label{eq:abm-choice}
\end{equation}
where $\operatorname{dist}$ is network distance, $z(\cdot)$ is the park zone (a $k$-means partition of the
attraction coordinates into six zones), and $\pi_\theta\ge1$ is a zone-persistence factor (wader $3.0$,
swimmer $1.7$, diver $1.0$) that biases visit sequencing toward staying in the current zone. The a priori
thrill weights encode the typology,
\begin{equation}
w_{\text{diver}}(\ell)=1+\tfrac12(\ell-1),\qquad
w_{\text{wader}}(\ell)=1+\tfrac12(5-\ell),\qquad
w_{\text{swimmer}}(\ell)=1+\tfrac12\big(2-|\ell-3|\big),
\label{eq:abm-weights}
\end{equation}
so divers tilt to high thrill, waders to low thrill, and swimmers to moderate rides (about a
three-to-one odds ratio across the range). These weights are set from the typology, not fit to the
observed heatmaps. The simulated field is built to mirror the observed heatmaps, which are a time-weighted
density of continuous GPS presence on the pathways (walking plus queuing). Accordingly each agent's
transit time is laid along the walked route, and its queue/dwell time
($\tau_{\text{attr}}=\SI{250}{\second}$ at attractions, $\tau_{\text{comm}}=\SI{120}{\second}$ at
commercial stops) is laid along the pathway approaching the destination (within \SI{25}{\metre}), not as
an off-path point. To reproduce the way real visitors spread across the network rather than collapsing
onto single shortest paths, every laid point carries a small lateral jitter (Gaussian, \SI{5}{\metre}) and
each leg draws one of eight near-optimal route variants (shortest paths on lognormally perturbed edge
weights). The field is then rasterized with the same kernel-density settings as the observed heatmaps.

We evaluate three populations on identical networks and rasterization: \emph{locomotion-only}
($w_\theta\equiv1$, $\pi_\theta\equiv1$; types differ only in $f_2$, $f_7$, $\gamma_\theta$),
\emph{full} (Eqs.~\ref{eq:abm-choice}--\ref{eq:abm-weights}), and \emph{homogeneous} (all types share the
mean parameters). For a field $\rho$ with pathway mean $\bar\rho$, write the relative occupancy
$\tilde\rho=\rho/\bar\rho$ and define the thrill-versus-commercial gradient of type $\theta$ as
\begin{equation}
G_\theta=\big\langle\tilde\rho_\theta\big\rangle_{\mathcal H}-\big\langle\tilde\rho_\theta\big\rangle_{\mathcal C},
\label{eq:abm-gradient}
\end{equation}
where $\mathcal H$ is the set of pixels within \SI{18}{\metre} of a level-$4$ or $5$ ride and $\mathcal C$
the commercial pixels. The primary statistic is the diver-minus-wader separation
$G_{\text{diver}}-G_{\text{wader}}$, reported with percentile $95\%$ intervals over $16$ random seeds;
we also note whether each model's profile $(G_{\text{wader}},G_{\text{swimmer}},G_{\text{diver}})$ rises
monotonically from waders to divers as the observed one does. As a secondary check we retain the pixelwise
simulated-versus-observed correlation (Table~\ref{tab:abm-matrix}).

\section{Supporting analyses}\label{sec:support}
\begin{figure}[H]
\centering
\includegraphics{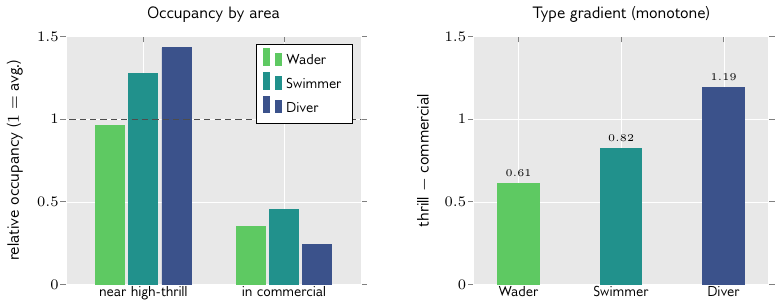}
\caption{Observed Knott's occupancy by visitor type, relative to the park average ($1=$ average),
measured directly from the cluster heatmaps. Left: occupancy near high-thrill rides (within
\SI{18}{\metre} of level 4--5 attractions) versus in commercial areas. Right: the resulting
thrill-minus-commercial gradient $G_\theta$ (Eq.~\ref{eq:abm-gradient}) increases monotonically from
waders to divers. This type signature is model-free; it corresponds to the ``Observed'' row of
Table~\ref{tab:abm-matrix}.}
\label{sfig:abm-signature}
\end{figure}

\begin{figure}[H]
\centering
\includegraphics{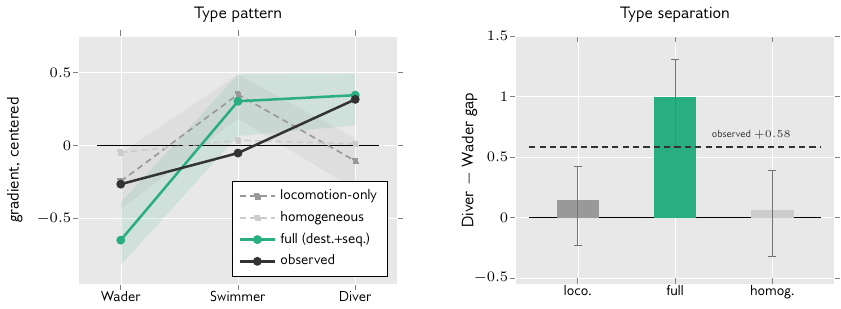}
\caption{Which model reproduces the observed gradient. Left: the thrill-minus-commercial gradient by
type, centered within each profile to isolate how the types differ; the full destination-choice and
sequencing model (green) tracks the observed shape (black), while the locomotion-only and homogeneous
models are flat. Right: the diver-minus-wader separation with \SI{95}{\percent} intervals over 16 seeds;
only the full model produces a separation excluding zero. It somewhat overstates the magnitude (the
a priori weights are uncalibrated) but reproduces the direction and ordering that locomotion parameters
alone cannot. This is the visual form of the three model rows of Table~\ref{tab:abm-matrix}.}
\label{sfig:abm-comparison}
\end{figure}

This section gives the full tables behind results the main text reports only in summary:
Table~\ref{tab:validity-full} reports cluster validity across $k=2$ to $5$;
Table~\ref{tab:reproducibility} the reproducibility measures (restart stability, cross-algorithm
agreement, and consensus); Table~\ref{tab:directionality-full} the full cross-park directionality;
Table~\ref{tab:abm-matrix} with Figures~\ref{sfig:abm-signature} and~\ref{sfig:abm-comparison} the
agent-based occupancy-gradient comparison across the three populations; and
Table~\ref{tab:survey-park} the per-park exit-survey aggregates. All values are produced by the analysis
scripts (see Data and code availability).

\begin{table}[H]
\centering
\caption{Cluster-validity indices for $k=2$ to $5$ at each park (canonical $k$-means). The main text
reports only $k=2$ and $3$. DCA admits at most $k=3$ at $n=4$. Higher silhouette,
Caliński--Harabasz, and gap, and lower Davies--Bouldin, indicate better-separated clusters.}
\label{tab:validity-full}
\small
\begin{tabular}{llcccc}
\toprule
Park & $k$ & Silhouette & Davies--Bouldin & Caliński--Harabasz & Gap \\
\midrule
\multirow{4}{*}{KBF ($n{=}15$)}
 & 2 & 0.26 & 1.23 & 6.57 & $0.10$ \\
 & 3 & 0.24 & 1.22 & 6.42 & $0.18$ \\
 & 4 & 0.23 & 1.09 & 6.12 & $0.22$ \\
 & 5 & 0.23 & 1.06 & 6.00 & $0.25$ \\
\midrule
\multirow{4}{*}{Disneyland ($n{=}7$)}
 & 2 & 0.32 & 0.44 & 3.65 & $-0.43$ \\
 & 3 & 0.12 & 0.92 & 3.21 & $-0.44$ \\
 & 4 & 0.10 & 0.75 & 3.07 & $-0.49$ \\
 & 5 & 0.08 & 0.56 & 3.18 & $-0.54$ \\
\midrule
\multirow{2}{*}{DCA ($n{=}4$)}
 & 2 & 0.23 & 0.43 & 2.68 & $-0.92$ \\
 & 3 & 0.06 & 0.44 & 2.27 & $-1.11$ \\
\bottomrule
\end{tabular}
\end{table}

\begin{table}[H]
\centering
\caption{Reproducibility of the $k=3$ partitions. Restart stability is the mean adjusted Rand index (ARI)
between the canonical solution and 300 random-initialization runs (the high DCA value is an artifact of
$n=4$ admitting essentially one partition). Cross-algorithm agreement and subsampling consensus were
computed for the coherent Knott's partition.}
\label{tab:reproducibility}
\small
\begin{tabular}{lc}
\toprule
Measure & Value \\
\midrule
Restart-stability ARI (KBF)        & $0.62 \pm 0.28$ \\
Restart-stability ARI (Disneyland) & $0.36 \pm 0.25$ \\
Restart-stability ARI (DCA)        & $0.89 \pm 0.35$ \\
\midrule
KBF: $k$-means vs.\ Ward (ARI)       & $1.00$ \\
KBF: $k$-means vs.\ Gaussian mixture (ARI) & $0.48$ \\
KBF: Ward vs.\ Gaussian mixture (ARI) & $0.48$ \\
\midrule
KBF consensus: mean within-cluster $C_{ij}$  & $0.79$ \\
KBF consensus: mean between-cluster $C_{ij}$ & $0.09$ \\
\bottomrule
\end{tabular}
\end{table}

\begin{table}[H]
\centering
\caption{Cross-park directionality: standardized (diver $-$ wader) difference $\delta_j$ per feature per
park, in standard-deviation units (Eq.~\ref{eq:direction}). Three features (marked $\dagger$) reverse
sign between Knott's and a Disney park.}
\label{tab:directionality-full}
\small
\begin{tabular}{lccc}
\toprule
Feature & KBF & Disneyland & DCA \\
\midrule
\% time on paths$^\dagger$        & $1.26$ & $-0.02$ & $-1.98$ \\
Average speed                     & $1.50$ & $1.33$  & $0.87$ \\
POIs per hour$^\dagger$           & $1.98$ & $-0.06$ & $-1.58$ \\
Attractions per hour              & $2.62$ & $0.25$  & $1.89$ \\
Average attraction duration$^\dagger$ & $-2.00$ & $0.68$ & $2.09$ \\
Path directness                   & $1.16$ & $1.84$  & $1.73$ \\
Attraction focus                  & $1.74$ & $0.25$  & $2.14$ \\
Time to first attraction          & $-1.39$ & $-1.55$ & $-1.48$ \\
\bottomrule
\end{tabular}
\end{table}

\begin{table}[H]
\centering
\caption{Agent-based model: thrill-versus-commercial occupancy gradient $G_\theta$ by visitor type
(Eq.~\ref{eq:abm-gradient}), relative to the park average, for the observed heatmaps and the three
simulated populations (mean over $16$ seeds; $95\%$ interval on the Diver$-$Wader separation). The
observed gradient is monotone (wader $<$ swimmer $<$ diver). The discriminating statistic is the
Diver$-$Wader separation: only the full destination-choice-and-sequencing model produces a separation that
excludes zero and a profile that rises monotonically from waders to divers; the locomotion-only profile
instead peaks at swimmers and the homogeneous profile is nearly flat, and both of their separation
intervals include zero. The full model somewhat overstates the magnitude ($+0.99$ vs.\ the observed
$+0.58$), as expected for a priori, uncalibrated weights. As a secondary check, the pixelwise
simulated-versus-observed contrast correlation is near zero for every model (mean diagonal $r$:
locomotion-only $-0.00$, full $+0.00$, homogeneous $-0.03$), confirming that the $n=3$--$7$-participant
heatmaps cannot be resolved pixel by pixel.}
\label{tab:abm-matrix}
\small
\begin{tabular}{lcccc}
\toprule
 & Wader & Swimmer & Diver & Diver$-$Wader \\
\midrule
Observed          & $+0.61$ & $+0.82$ & $+1.19$ & $+0.58$ \\
Locomotion-only   & $+0.89$ & $+1.48$ & $+1.03$ & $+0.14$ \,[$-0.23,\,+0.42$] \\
Full (dest.+seq.) & $+0.56$ & $+1.51$ & $+1.55$ & $+0.99$ \,[$+0.58,\,+1.30$] \\
Homogeneous       & $+1.04$ & $+1.13$ & $+1.10$ & $+0.06$ \,[$-0.32,\,+0.39$] \\
\bottomrule
\end{tabular}
\end{table}

\begin{table}[H]
\centering
\caption{Exit-survey aggregates by park (all participants). Planning score is on $0$ to $2$;
self-description is on $0$ (wandering) to $1$ (beelining); map consultation is the fraction of
participants who consulted a map. These park-level aggregates complement the main text's KBF
cluster-level survey table.}
\label{tab:survey-park}
\small
\begin{tabular}{lcccc}
\toprule
Park & Planning score & Months since last visit (median) & Self-description & \% consulted map \\
\midrule
KBF        & 1.01 & 3.17 (1.0) & 0.31 & 48.3 \\
Disneyland & 0.96 & 4.90 (3.0) & 0.63 & 42.9 \\
DCA        & 1.08 & 6.67 (7.5) & 0.83 & 50.0 \\
\bottomrule
\end{tabular}
\end{table}

The Knott's feature-correlation structure (the four pairs with $|\rho|\ge0.6$: POIs/h with attractions/h,
$\rho=0.66$; POIs/h with \% time on paths, $0.65$; attractions/h with attraction focus, $0.65$; and
attractions/h with attraction duration, $-0.61$) is shown in full as a figure in the main text and is not
duplicated here.

\section{Two-cluster ($k=2$) comparison}\label{sec:k2}
\begin{figure}[H]
\centering
\includegraphics{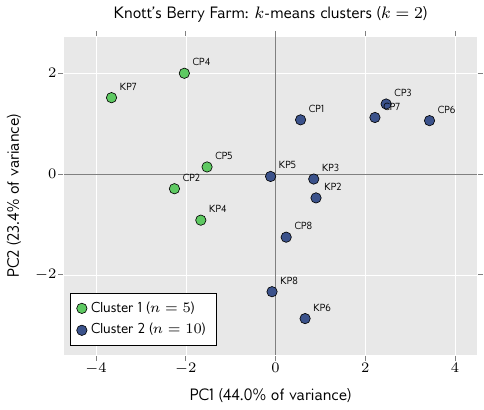}\hfill
\includegraphics{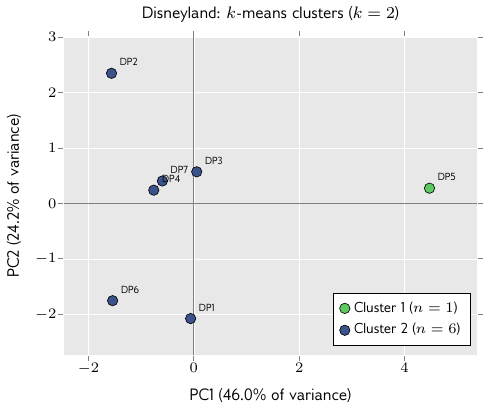}\\[2pt]
\includegraphics{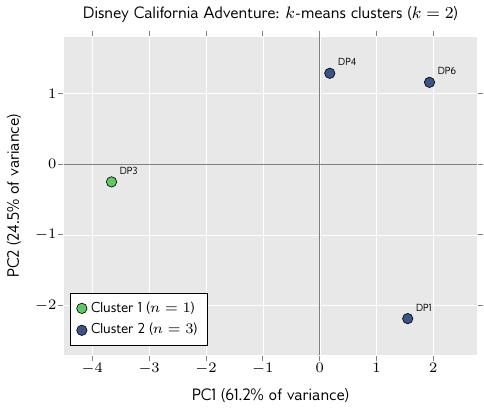}
\caption{Two-cluster $k$-means in principal-component space for Knott's Berry Farm, Disneyland, and DCA.}
\label{fig:k2}
\end{figure}

Following \textcite{Beeco2013GPS}, who contrasted ``wanderers'' and ``planners,'' we also ran $k=2$ at each
park. As Table~\ref{tab:validity-full} shows, the $k=2$ silhouette is marginally higher than $k=3$ at KBF
($0.26$ vs.\ $0.24$) and at the Disney parks, consistent with no strong natural number of clusters. We
retain $k=3$ in the main analysis because it corresponds to the wader/swimmer/diver typology under study;
the $k=2$ partitions in Figure~\ref{fig:k2} are shown only for comparison.

\section{Under-powered Disneyland and DCA analyses}\label{sec:dldca}
\begin{figure}[H]
\centering
\includegraphics{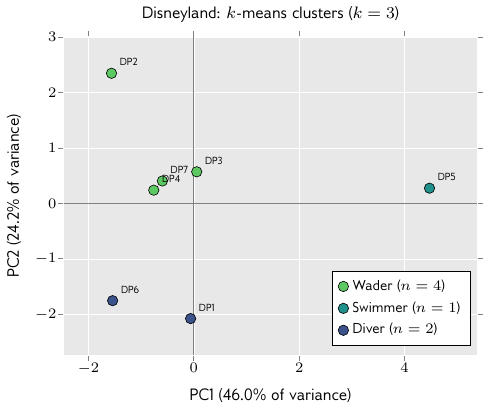}\hfill
\includegraphics{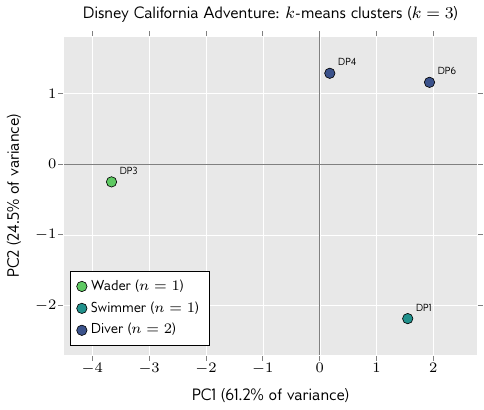}
\caption{Three-cluster $k$-means for Disneyland (left) and DCA (right). Cluster separation is weak and
unstable at these sample sizes.}
\label{fig:dldca-k3}
\end{figure}

\begin{figure}[H]
\centering
\includegraphics[width=\linewidth]{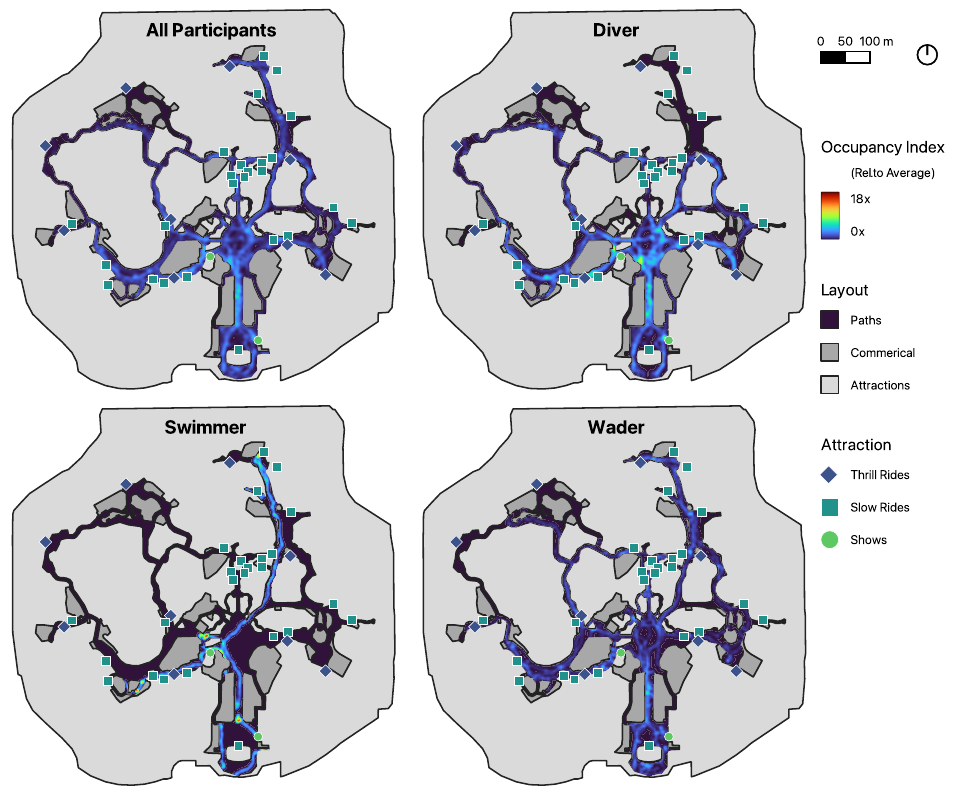}
\caption{Disneyland per-cluster occupancy heatmaps (exploratory; one participant in the swimmer cluster).
Contrast between clusters is muted relative to Knott's.}
\label{fig:heatmap-dl}
\end{figure}

\begin{figure}[H]
\centering
\includegraphics[width=\linewidth]{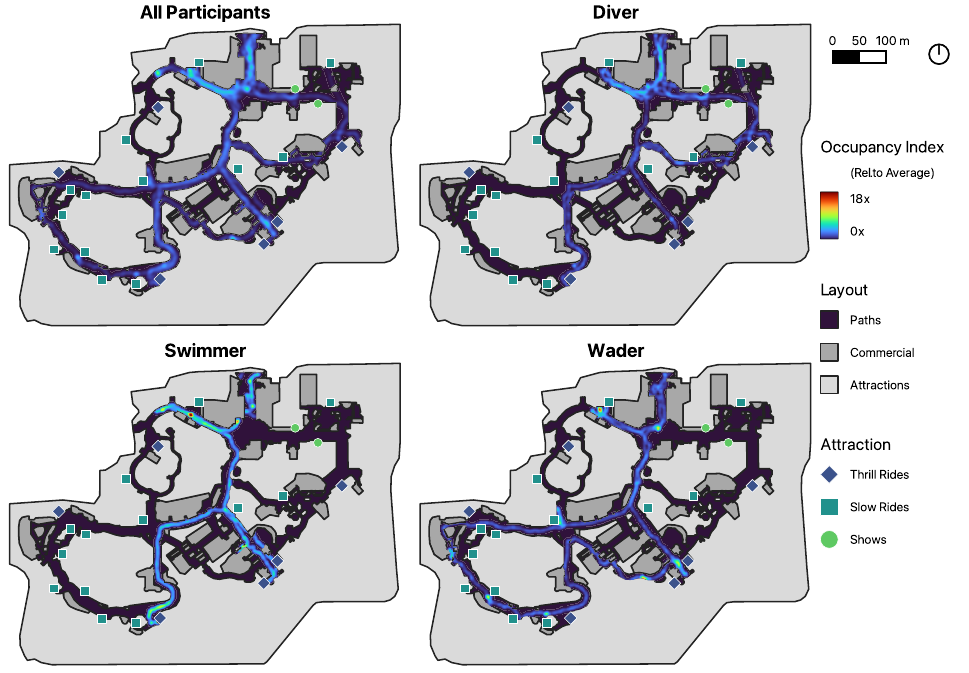}
\caption{DCA per-cluster occupancy heatmaps (exploratory; roughly one participant per cluster). Shown as
path visualizations only.}
\label{fig:heatmap-dca}
\end{figure}

With $n=7$ (Disneyland) and $n=4$ (DCA), neither park clears the a priori coherence threshold
(Table~\ref{tab:validity-full}): $k=3$ silhouettes are $0.12$ and $0.06$ respectively, with negative gap
statistics. We therefore exclude both from quantitative parameter recommendations and report their $k=3$
projections (Figure~\ref{fig:dldca-k3}), rank-average tables (Tables~\ref{tab:rank-dl}
and~\ref{tab:rank-dca}), and occupancy heatmaps (Figures~\ref{fig:heatmap-dl} and~\ref{fig:heatmap-dca})
here for completeness only. No hypothesis tests are performed at these sample sizes.

\begin{table}[H]
\centering
\begin{minipage}[t]{0.46\linewidth}\centering
\caption{Disneyland rank-average by participant ($k=3$).}
\label{tab:rank-dl}
\small
\begin{tabular}{llc}
\toprule
ID & Cluster & $R_i$ \\
\midrule
DP2 & Wader   & 0.339 \\
DP4 & Wader   & 0.375 \\
DP3 & Wader   & 0.554 \\
DP7 & Wader   & 0.589 \\
DP5 & Swimmer & 0.482 \\
DP1 & Diver   & 0.607 \\
DP6 & Diver   & 0.679 \\
\bottomrule
\end{tabular}
\end{minipage}\hfill
\begin{minipage}[t]{0.46\linewidth}\centering
\caption{DCA rank-average by participant ($k=3$).}
\label{tab:rank-dca}
\small
\begin{tabular}{llc}
\toprule
ID & Cluster & $R_i$ \\
\midrule
DP3 & Wader   & 0.406 \\
DP1 & Swimmer & 0.562 \\
DP4 & Diver   & 0.531 \\
DP6 & Diver   & 0.625 \\
\bottomrule
\end{tabular}
\end{minipage}
\end{table}

\section{Pooled cross-park model}\label{sec:pooled}
\begin{figure}[H]
\centering
\includegraphics{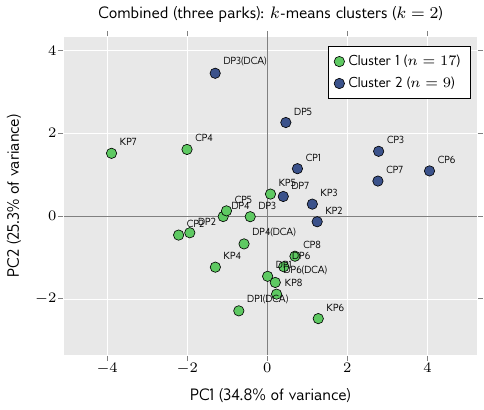}\hfill
\includegraphics{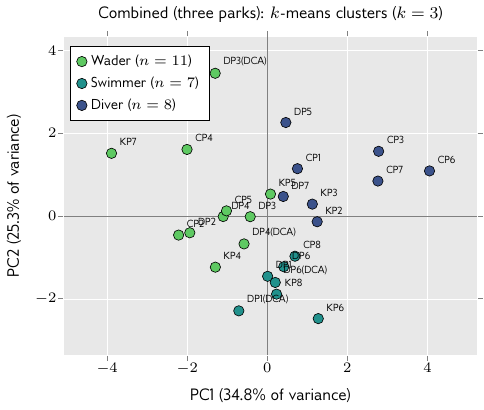}
\caption{Pooled cross-park $k$-means at $k=2$ (left) and $k=3$ (right). Participant identifiers suffixed
``(DCA)'' denote a park-hopper's DCA visit.}
\label{fig:combined}
\end{figure}

Pooling all three parks and clustering jointly yields incoherent assignments: participants who are clear
divers within their own park scatter across the pooled clusters (Figure~\ref{fig:combined};
Table~\ref{tab:rank-combined}), because feature directionality differs by park
(Table~\ref{tab:directionality-full}). This is presented as evidence \emph{for} park-specific modeling,
not as an alternative analysis.

\begin{table}[H]
\centering
\caption{Pooled cross-park rank-average by participant ($k=3$). Labels reflect the pooled clustering, not
the per-park clustering; several within-park divers are reassigned, illustrating why a pooled model is
inappropriate.}
\label{tab:rank-combined}
\small
\begin{tabular}{llc@{\hspace{3em}}llc}
\toprule
ID & Cluster & $R_i$ & ID & Cluster & $R_i$ \\
\midrule
KP7      & Wader   & 0.197 & DP6(DCA) & Swimmer & 0.579 \\
CP2      & Wader   & 0.219 & KP8      & Swimmer & 0.579 \\
DP3(DCA) & Wader   & 0.286 & CP8      & Swimmer & 0.596 \\
DP4      & Wader   & 0.308 & DP6      & Swimmer & 0.606 \\
DP2      & Wader   & 0.312 & KP6      & Swimmer & 0.743 \\
CP4      & Wader   & 0.320 & DP5      & Diver   & 0.498 \\
KP4      & Wader   & 0.385 & CP1      & Diver   & 0.531 \\
CP5      & Wader   & 0.466 & DP7      & Diver   & 0.543 \\
DP4(DCA) & Wader   & 0.478 & KP3      & Diver   & 0.603 \\
KP5      & Wader   & 0.490 & KP2      & Diver   & 0.668 \\
DP3      & Wader   & 0.510 & CP7      & Diver   & 0.680 \\
DP1(DCA) & Swimmer & 0.510 & CP3      & Diver   & 0.683 \\
DP1      & Swimmer & 0.572 & CP6      & Diver   & 0.762 \\
\bottomrule
\end{tabular}
\end{table}

\section{Per-park rank-average table (Knott's Berry Farm)}\label{sec:rank}
The rank-average (Eq.~\ref{eq:rankavg}) is used \emph{only} to order clusters for labeling, not as
validation; independent validation is via Kruskal--Wallis and the exit survey in the main text.
Table~\ref{tab:rank-kbf} lists the per-participant values for the coherent Knott's partition. Because
labels follow the cluster a participant belongs to, individual values can overlap across labels (for
example KP6, a diving-leaning swimmer).

\begin{table}[H]
\centering
\caption{Knott's rank-average by participant ($k=3$).}
\label{tab:rank-kbf}
\small
\begin{tabular}{llc}
\toprule
ID & Cluster & $R_i$ \\
\midrule
KP7 & Wader   & 0.200 \\
CP2 & Wader   & 0.250 \\
CP4 & Wader   & 0.317 \\
KP4 & Wader   & 0.388 \\
CP5 & Wader   & 0.442 \\
KP5 & Swimmer & 0.442 \\
CP1 & Swimmer & 0.496 \\
KP3 & Swimmer & 0.537 \\
CP8 & Swimmer & 0.554 \\
KP8 & Swimmer & 0.583 \\
KP2 & Swimmer & 0.633 \\
KP6 & Swimmer & 0.712 \\
CP7 & Diver   & 0.650 \\
CP3 & Diver   & 0.658 \\
CP6 & Diver   & 0.763 \\
\bottomrule
\end{tabular}
\end{table}

\section{Study instruments}\label{sec:instruments}
\ifdefined\anonymous
The recruitment script, consent form, and exit-survey questionnaire are omitted from this anonymous
version because the original documents carry institutional and author identifiers (letterhead, names,
contact details, and the IRB protocol number). They are reproduced in full in the version with author
details and are available to the editors on request.
\else
The IRB-approved recruitment script (Figure~\ref{fig:recruit}), consent form (Figure~\ref{fig:consent}),
and exit-survey questionnaire (Figure~\ref{fig:survey}) are reproduced below.

\begin{figure}[H]
\centering
\includegraphics[height=0.85\textheight,keepaspectratio]{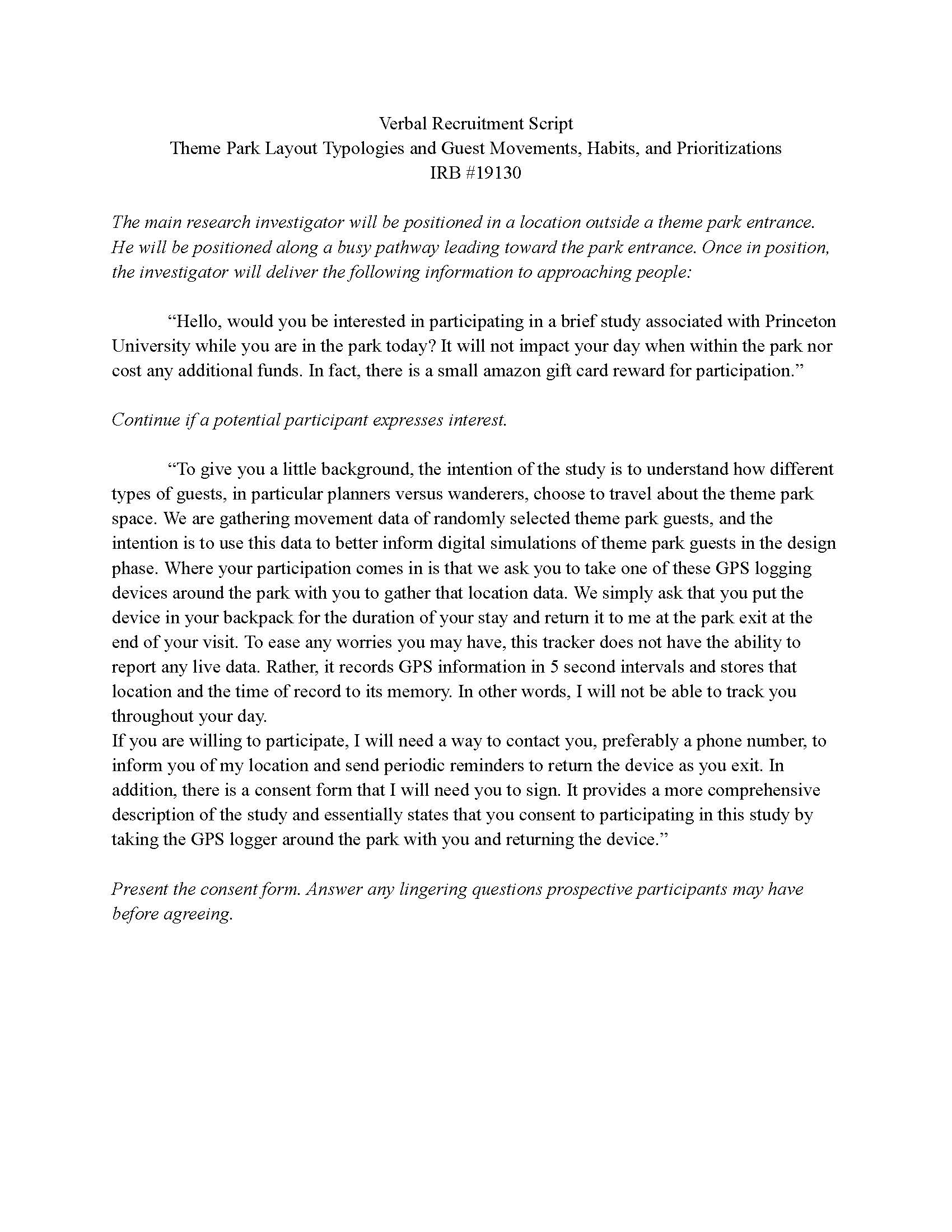}
\caption{Verbal recruitment script (Princeton University IRB \#19130).}
\label{fig:recruit}
\end{figure}

\begin{figure}[H]
\centering
\includegraphics[width=0.49\linewidth]{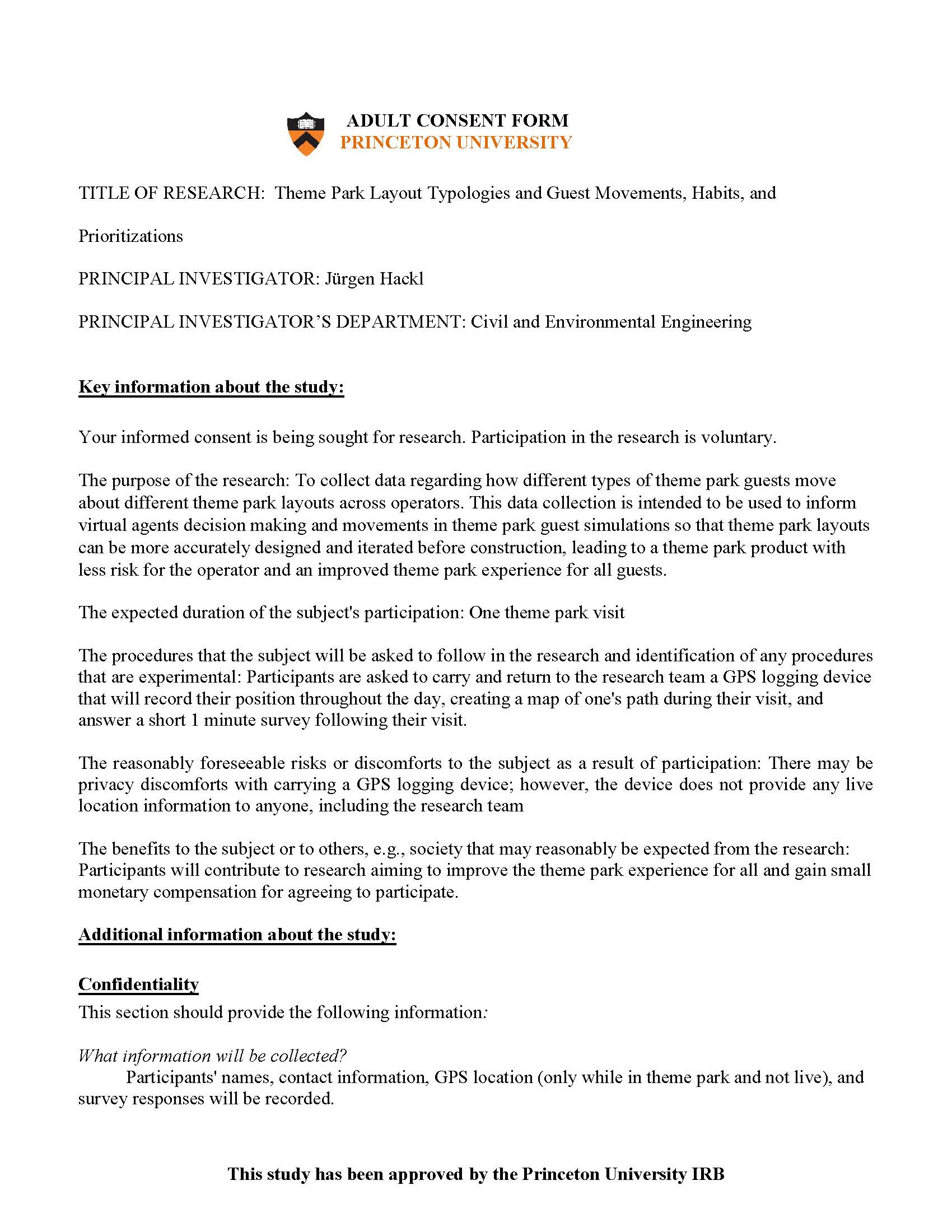}
\includegraphics[width=0.49\linewidth]{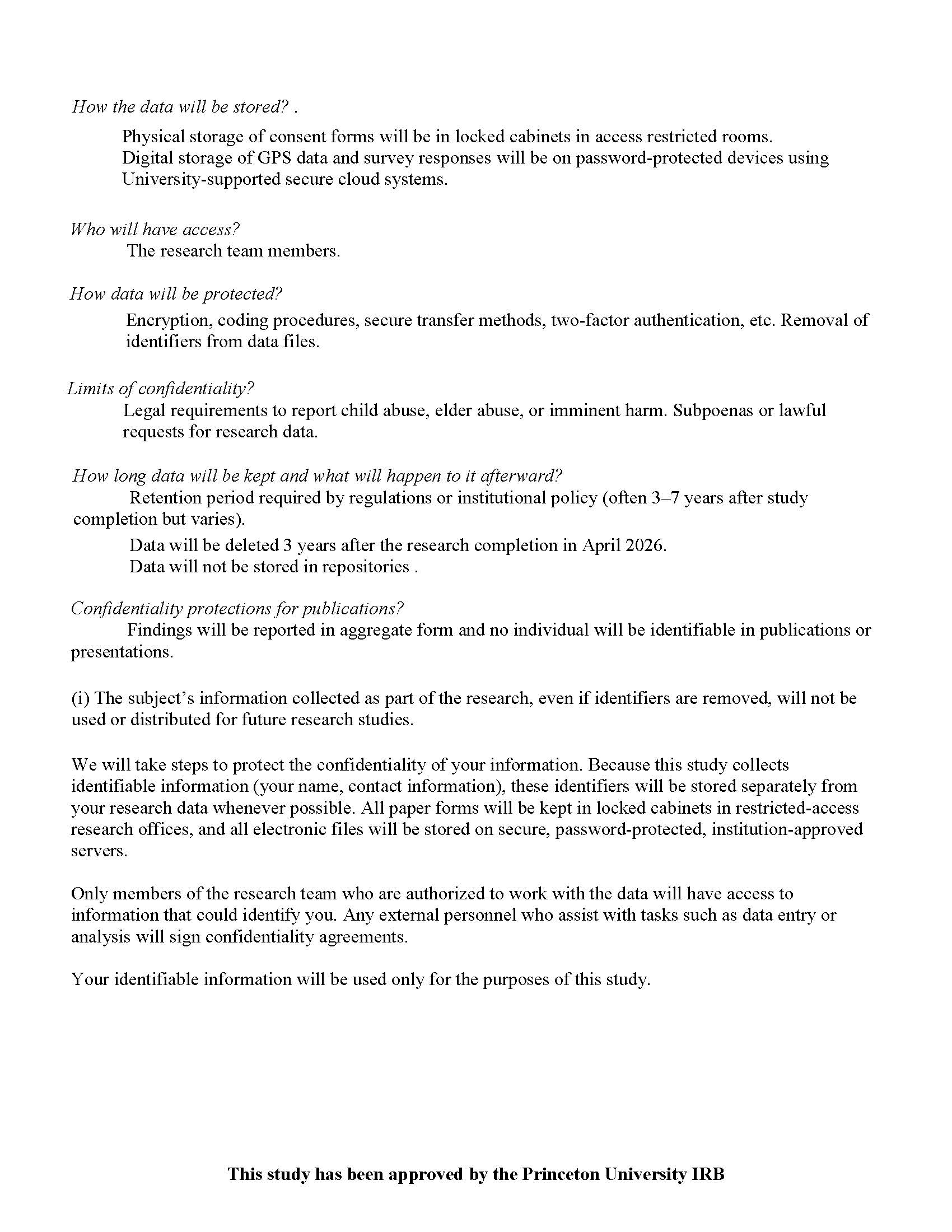}\\[2pt]
\includegraphics[width=0.49\linewidth]{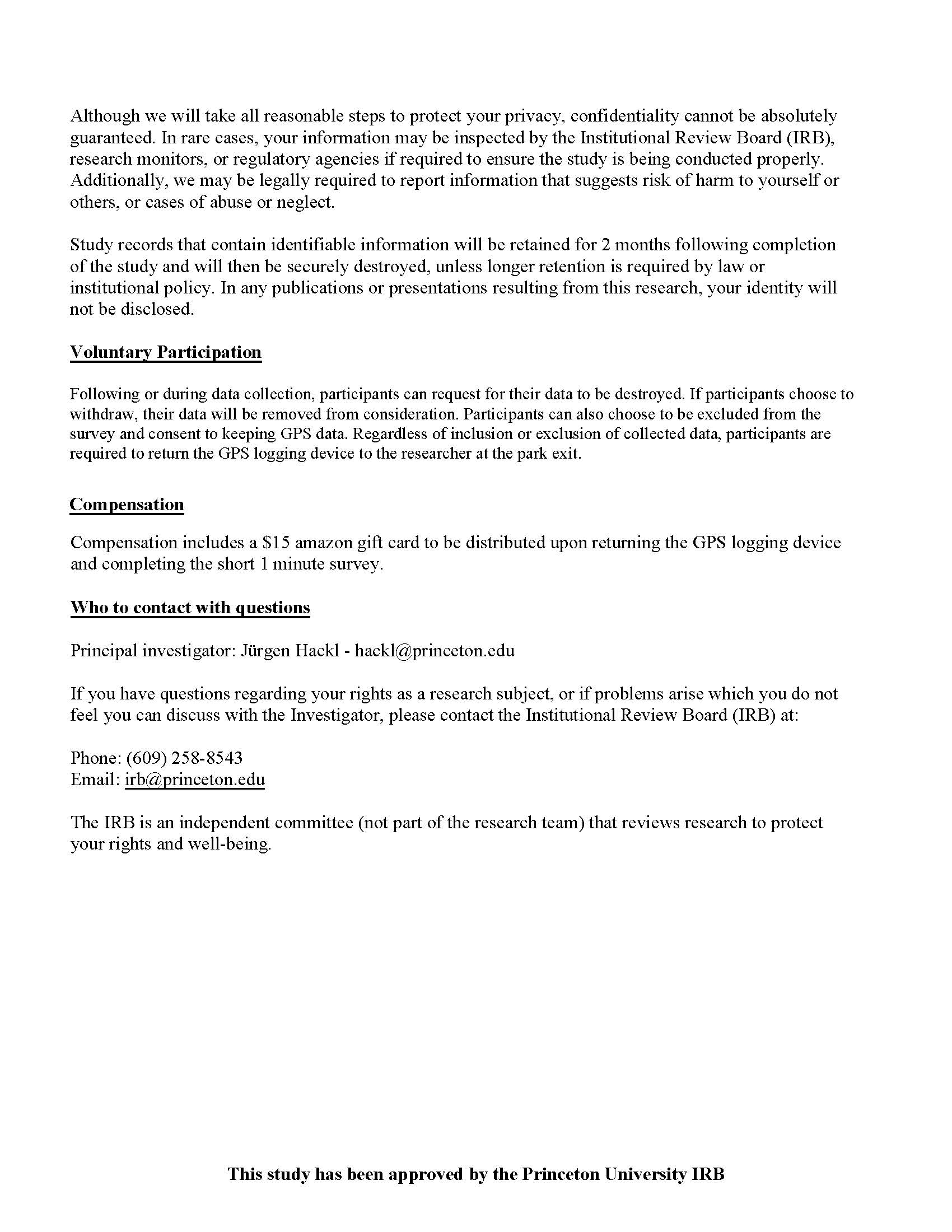}
\includegraphics[width=0.49\linewidth]{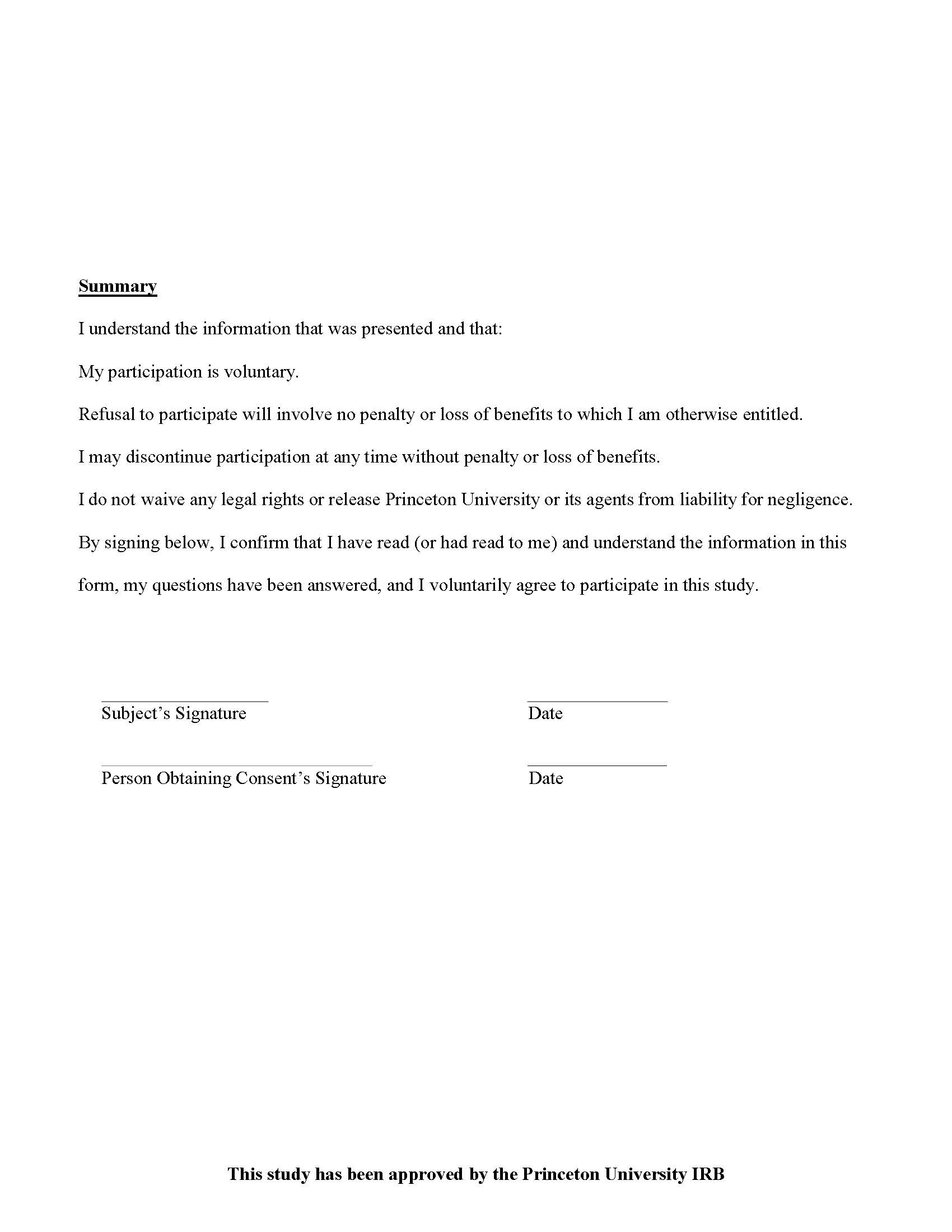}
\caption{Participant consent form (four pages).}
\label{fig:consent}
\end{figure}

\begin{figure}[H]
\centering
\includegraphics[height=0.85\textheight,keepaspectratio]{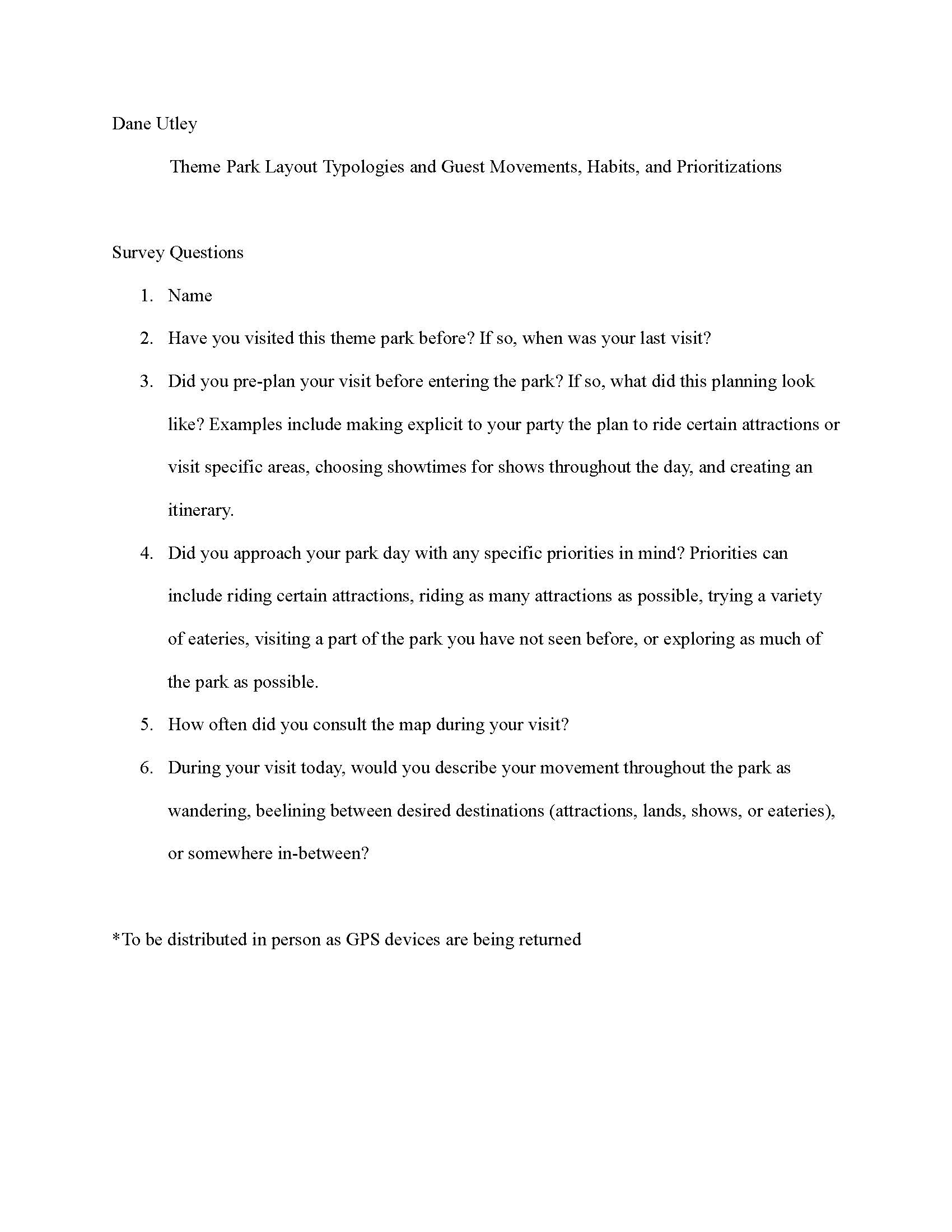}
\caption{Exit survey questionnaire.}
\label{fig:survey}
\end{figure}
\fi

\section*{Data and code availability}
\ifdefined\anonymous
The de-identified data, together with the analysis and agent-based-simulation code that regenerates every
main-text and supplementary figure, table, and statistic, are archived in a single public repository under
open licenses (CC-BY-4.0 for the data, MIT for the code). The repository citation is withheld here to
preserve anonymity and will be provided upon acceptance.
\else
The de-identified data are archived in a single Zenodo deposit
(DOI:~\href{https://doi.org/10.5281/zenodo.20748587}{10.5281/zenodo.20748587}); the
same deposit includes, under \texttt{code/}, the analysis and agent-based-simulation code that regenerates
every main-text and supplementary figure, table, and statistic directly from the deposited data. The data
are released under CC-BY-4.0 and the code under the MIT License.
\fi

\printbibliography[title={Supplementary References}]
\end{refsection}

\end{document}